\documentclass[zpreprint,zbstpl]{zeus_paper}
\usepackage[english]{babel}
\chardef\usc=95
\chardef\til=126
\catcode`\@=11 
\DeclareRobustCommand\xdotspace{\futurelet\@let@token\@xdotspace}
\def\@xdotspace{%
  \ifx\@let@token.\else
  \ifx\@let@token\bgroup.\else
  \ifx\@let@token\egroup.\else
  \ifx\@let@token\/.\else
  \ifx\@let@token\ .\else
  \ifx\@let@token~.\else
  \ifx\@let@token!.\else
  \ifx\@let@token,.\else
  \ifx\@let@token:.\else
  \ifx\@let@token;.\else
  \ifx\@let@token?.\else
  \ifx\@let@token/.\else
  \ifx\@let@token'.\else
  \ifx\@let@token).\else
  \ifx\@let@token-.\else
  \ifx\@let@token\@xobeysp.\else
  \ifx\@let@token\space.\else
  \ifx\@let@token\@sptoken.\else
   .\space
   \fi\fi\fi\fi\fi\fi\fi\fi\fi\fi\fi\fi\fi\fi\fi\fi\fi\fi}
\catcode`\@=12 

\newcommand{\stru}[2]{%
   \relax\ifmmode\hbox{\vrule height#1 depth#2 width0pt}%
   \else\vrule height#1 depth#2 width0pt\fi}

\newcommand{\Ronum}[1]{\uppercase\expandafter{\romannumeral#1}}
\newcommand{\ronum}[1]{\expandafter{\romannumeral#1}}
\DeclareRobustCommand{\LaTeXZ}{%
  \LaTeX\kern-.05em4\kern-.1em
  {\raisebox{-0.2ex}{$\scriptstyle\text{ZEUS}$}}\xspace}

\DeclareMathAlphabet{\mathbf}{OT1}{cmr}{bx}{sl}
\newcommand{\eVdist}{\kern-0.06667em}

\newcommand{\Gev}{{\text{Ge}\eVdist\text{V\/}}}

\newcommand{\mev}{{\,\text{Me}\eVdist\text{V\/}}}
\newcommand{\gev}{{\,\text{Ge}\eVdist\text{V\/}}}

\newcommand{\nb}{\,\text{nb}}
\newcommand{\pb}{\,\text{pb}}

\newcommand{\slashfrac}[2]{%
  \raisebox{0.5ex}{\ensuremath #1}\kern-0.12em/\kern-0.08em
  \raisebox{-.8ex}{\ensuremath #2}}

\newcommand{\sqr}[3]{%
    {\vcenter{\hrule height.#3ex\hbox{\vrule width.#2ex height#1ex
     \kern#1ex\vrule width.#3ex}\hrule height.#2ex}}}

\catcode`\@=11 
\newcommand{\parenbar}{\mathpalette\p@renb@r}
\def\p@renb@r#1#2{\vbox{%
  \ifx#1\scriptscriptstyle \dimen@.7em\dimen@ii.2em\else
  \ifx#1\scriptstyle \dimen@.8em\dimen@ii.25em\else
  \dimen@1em\dimen@ii.4em\fi\fi \offinterlineskip
  \ialign{\hfill##\hfill\cr
    \vbox{\hrule width\dimen@ii}\cr
    \noalign{\vskip-.3ex}%
    \hbox to\dimen@{$\mathchar300\hfil\mathchar301$}\cr
    \noalign{\vskip-.3ex}%
    $#1#2$\cr}}}
\catcode`\@=12 

\newcommand{\IP}{{\rm I$\kern-0.01667em$P}\xspace}
\newcommand{\JB}{{\rm JB}}

\mathchardef\qsm=63
\mathchardef\pls=43
\mathchardef\mns=512
\mathchardef\plm=518
\mathchardef\eql=61
\mathchardef\smallleft=300
\mathchardef\smallright=301
\mathchardef\les=316
\mathchardef\gre=318
\mathchardef\leq=532
\mathchardef\grq=533

\catcode`\@=11 
\newcounter{pict@width}
\newcounter{pict@height}
\newlength{\pict@scale}
\newcommand{\psfigadd}[4]{%
\setcounter{pict@width}{1*\ratio{#2+\pict@scale/2}{\pict@scale}}
\setcounter{pict@height}{1*\ratio{#3+\pict@scale/2}{\pict@scale}}
\setlength{\unitlength}{\pict@scale}
\hbox to #2{\hspace{-\fill}\begin{picture}(\thepict@width,\thepict@height)
\put(0,0){\psfig{figure=#1,width=#2,height=#3,clip=}}
\SetScale{0.283466457}
\SetWidth{1.763889}
{#4}
\end{picture}}
}
\newcounter{pict@widthfst}
\newcounter{pict@widthscd}
\newcounter{pict@widthtot}
\newcommand{\psfigaddtwo}[7]{%
\setcounter{pict@widthfst}{1*\ratio{#2+\pict@scale/2}{\pict@scale}}
\setcounter{pict@widthscd}{1*\ratio{#2+#4+\pict@scale/2}{\pict@scale}}
\setcounter{pict@widthtot}{1*\ratio{#2+#4+#6+\pict@scale/2}{\pict@scale}}
\setcounter{pict@height}{1*\ratio{#3+\pict@scale/2}{\pict@scale}}
\setlength{\unitlength}{\pict@scale}
\hbox{\hspace{-\fill}\begin{picture}(\thepict@widthtot,\thepict@height)
\put(0,0){\psfig{figure=#1,width=#2,height=#3,clip=}}
\put(\thepict@widthscd,0){\psfig{figure=#5,width=#6,height=#3,clip=}}
\SetScale{0.283466457}
\SetWidth{1.763889}
{#7}
\end{picture}}
}
\newcommand{\psfigror}[4]{%
\setcounter{pict@width}{1*\ratio{#2+\pict@scale/2}{\pict@scale}}
\setcounter{pict@height}{1*\ratio{#3+\pict@scale/2}{\pict@scale}}
\setlength{\unitlength}{\pict@scale}
\hbox{\begin{picture}(\thepict@width,\thepict@height)
\put(0,\thepict@height){\psfig{figure=#1,width=#3,height=#2,clip=,angle=270}}
\SetScale{0.283466457}
\SetWidth{1.763889}
{#4}
\end{picture}}
}
\newcommand{\psfigrol}[4]{%
\setcounter{pict@width}{1*\ratio{#2+\pict@scale/2}{\pict@scale}}
\setcounter{pict@height}{1*\ratio{#3+\pict@scale/2}{\pict@scale}}
\setlength{\unitlength}{\pict@scale}
\hbox{\begin{picture}(\thepict@width,\thepict@height)
\put(0,0){\psfig{figure=#1,width=#3,height=#2,clip=,angle=90}}
\SetScale{0.283466457}
\SetWidth{1.763889}
{#4}
\end{picture}}
}
\catcode`\@=12 
\newlength\listtextwidth

\newcommand{\pcite}[1]{{\protect\cite{#1}}}

\catcode`\@=11 
\newlength{\@tabfninsert}
\newlength{\@tabfnwidth}
\newcommand{\tabfootnote}[2]{%
  \setlength{\@tabfninsert}{0.8em}
  \setlength{\@tabfnwidth}{\textwidth}
  \addtolength{\@tabfnwidth}{-\@tabfninsert}
  \addtolength{\@tabfnwidth}{-0.4em}
  \noindent\makebox[\@tabfninsert][r]{\footnotesize$^{#1}$\hfil}\hfill%
  \parbox[t]{\@tabfnwidth}{\footnotesize #2\hfill}}
\catcode`\@=12 

\newcommand{\GeVs}      {\mbox{${\,\rm GeV}^2$}}
\newcommand{\GeV}       {\mbox{${\,\rm GeV}$}}

\newcommand {\pom} {I\!\!P}
\newcommand {\pomsub} {{\scriptscriptstyle \pom}}

\newcommand{\dsp}        {\mbox{$D^{\ast +}$}}

\newcommand{\dz}         {\mbox{$D^{0}$}}
\def\dsk3pi{ {\dsp}~\rightarrow~\dz~\pi^{+}_{s}%
        \rightarrow~(K^{-}~\pi^{+}~\pi^{+}~\pi^{-})~\pi^{+}_{s} }
\def\et10t{ E_T^{\theta > 10^\circ}}
\begin{document}
%
%
%
\prepnum{DESY--07--039}

\title{
Diffractive  Photoproduction of~ \boldmath $D^{*\pm}(2010)$ at HERA
} 
                    
\author{ZEUS Collaboration}
\date{March 2007}

\abstract{
      Diffractive photoproduction of $D^{*\pm}(2010)$ mesons 
      was measured  
      with the ZEUS detector at the $ep$ collider HERA,
      using an integrated luminosity of $78.6\pb^{-1}$.  
      The $D^*$ mesons were reconstructed in the kinematic range:
      transverse momentum $p_T(D^*) > 1.9 \gev$ and   
      pseudorapidity  $\vert \eta(D^*)\vert < 1.6$, 
      using the decay $D^{*+} \to D^0 \pi^+_s$ 
      followed by $D^0 \to K^- \pi^+$ (+c.c.).     
      Diffractive events were identified by a large gap in pseudorapidity 
      between the produced hadronic state and the outgoing proton.
      Cross sections are reported for 
      photon-proton centre-of-mass energies in the range 
      $130 < W < 300 \gev$ ~and for 
      photon virtualities $Q^2 < 1 \gev^2$,
      in two ranges of the Pomeron fractional momentum 
      $ x_{\pom} < 0.035$ and $ x_{\pom} < 0.01$. 
      The relative contribution of diffractive events to the  
      inclusive $D^{*\pm}(2010)$ photoproduction  cross section 
      is about 6\%. The data are in agreement with
      perturbative QCD calculations based on 
      various parameterisations of diffractive 
      parton distribution functions.  
      The  results are consistent with 
      diffractive QCD factorisation. 
}

\makezeustitle

\def\3{\ss}
\pagenumbering{Roman}
\begin{center}                                                                                     
{                      \Large  The ZEUS Collaboration              }                               
\end{center}                                                                                       
  S.~Chekanov$^{   1}$,                                                                            
  M.~Derrick,                                                                                      
  S.~Magill,                                                                                       
  B.~Musgrave,                                                                                     
  D.~Nicholass$^{   2}$,                                                                           
  \mbox{J.~Repond},                                                                                
  R.~Yoshida\\                                                                                     
 {\it Argonne National Laboratory, Argonne, Illinois 60439-4815}, USA~$^{n}$                       
\par \filbreak                                                                                     
  M.C.K.~Mattingly \\                                                                              
 {\it Andrews University, Berrien Springs, Michigan 49104-0380}, USA                               
\par \filbreak                                                                                     
  M.~Jechow, N.~Pavel~$^{\dagger}$, A.G.~Yag\"ues Molina \\                                        
  {\it Institut f\"ur Physik der Humboldt-Universit\"at zu Berlin,                                 
           Berlin, Germany}
% ...
\newpage                                                                                           
$^{\    1}$ supported by DESY, Germany \\                                                          
$^{\    2}$ also affiliated with University College London, UK \\                                  
% ...
$^{\dagger}$ deceased \\                                                                           
%
\newpage   
                                                           %
                                                           %
\begin{tabular}[h]{rp{14cm}}                                                                       
$^{a}$ &  supported by the Natural Sciences and Engineering Research Council of Canada (NSERC) \\  
$^{b}$ &  supported by the German Federal Ministry for Education and Research (BMBF), under        
          contract numbers HZ1GUA 2, HZ1GUB 0, HZ1PDA 5, HZ1VFA 5\\                                
$^{c}$ &  supported in part by the MINERVA Gesellschaft f\"ur Forschung GmbH, the Israel Science   
          Foundation (grant no. 293/02-11.2) and the U.S.-Israel Binational Science Foundation \\  
$^{d}$ &  supported by the German-Israeli Foundation and the Israel Science Foundation\\           
$^{e}$ &  supported by the Italian National Institute for Nuclear Physics (INFN) \\                
$^{f}$ &  supported by the Japanese Ministry of Education, Culture, Sports, Science and Technology 
          (MEXT) and its grants for Scientific Research\\                                          
$^{g}$ &  supported by the Korean Ministry of Education and Korea Science and Engineering          
          Foundation\\                                                                             
$^{h}$ &  supported by the Netherlands Foundation for Research on Matter (FOM)\\                   
$^{i}$ &  supported by the Polish State Committee for Scientific Research, grant no.               
          620/E-77/SPB/DESY/P-03/DZ 117/2003-2005 and grant no. 1P03B07427/2004-2006\\             
$^{j}$ &  partially supported by the German Federal Ministry for Education and Research (BMBF)\\   
$^{k}$ &  supported by RF Presidential grant N 8122.2006.2 for the leading                         
          scientific schools and by the Russian Ministry of Education and Science through its grant
          Research on High Energy Physics\\                                                        
$^{l}$ &  supported by the Spanish Ministry of Education and Science through funds provided by     
          CICYT\\                                                                                  
$^{m}$ &  supported by the Particle Physics and Astronomy Research Council, UK\\                   
$^{n}$ &  supported by the US Department of Energy\\                                               
$^{o}$ &  supported by the US National Science Foundation. Any opinion,                            
findings and conclusions or recommendations expressed in this material                             
are those of the authors and do not necessarily reflect the views of the                           
National Science Foundation.\\                                                                     
$^{p}$ &  supported by the Polish Ministry of Science and Higher Education                         
as a scientific project (2006-2008)\\                                                              
$^{q}$ &  supported by FNRS and its associated funds (IISN and FRIA) and by an Inter-University    
          Attraction Poles Programme subsidised by the Belgian Federal Science Policy Office\\     
$^{r}$ &  supported by the Malaysian Ministry of Science, Technology and                           
Innovation/Akademi Sains Malaysia grant SAGA 66-02-03-0048\\                                       
\end{tabular}                                                                                      
                                                           %
                                                           %
\clearpage

\pagenumbering{arabic} 
\pagestyle{plain}
%
%
\section{Introduction}
\label{sec-int}
%

In ~diffractive ~electron-proton~  scattering,  
the proton ~loses~ a small fraction ~of~ its energy and 
either ~emerges~ from the scattering ~intact, $ep \to eXp$, 
or ~dissociates~ into a low-mass state $N$, $ep \to eXN$.
A large gap in rapidity 
separates the hadronic-state $X$ with invariant-mass $M_X$
and the final-state proton (or $N$).

In  the framework of Regge phenomenology~\cite{ReggeTheory},
diffractive interactions  are ascribed to the exchange of
a trajectory with  vacuum quantum numbers, 
the Pomeron trajectory. In quantum chromodynamics (QCD), 
the diffractive factorisation theorem 
~\cite{trentadue,qcdf,*qcdf1,berera,QCDTheory}
states that the diffractive cross section,
in the presence of a hard scale, can be expressed as 
the convolution of universal partonic cross sections and 
a specific type of parton distribution function (PDF),
 the diffractive PDF (dPDF).  Diffractive PDFs are 
interpreted as conditional probabilities to find a
parton in the proton when the final state  
contains a fast forward proton. 
  The dPDFs   \cite{H1Fit2006,ZEUSLPS,GLP, MRW06} 
  have been determined
 from the HERA inclusive measurements of the  
 diffractive structure function ($F^D_2$), defined 
 in analogy with 
the proton structure function ($F_2$)  \cite{HERAdPDFs}, 
 and can be used as input 
  for  calculations of different diffractive processes, 
  for example at the Tevatron  and LHC~\cite{LHCTevatronDPDFS}. 
  
\begin{figure}[hbtp]
\epsfysize=12.5cm
\centerline{\epsffile{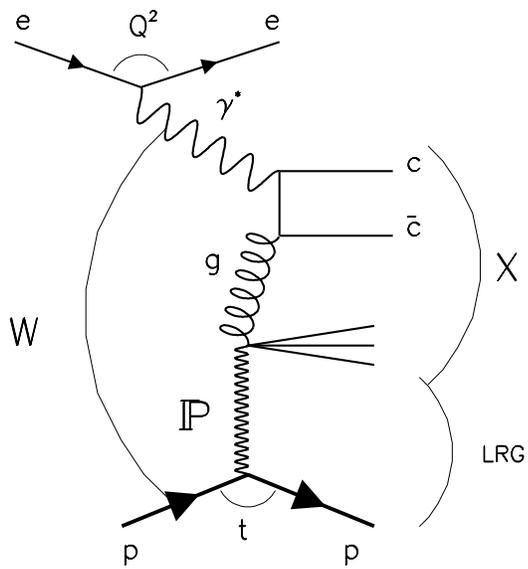}}
\vspace*{-2.0cm}
\caption{
Example of charm production in diffractive $ep$ scattering:
boson-gluon fusion in the resolved-Pomeron model \pcite{IS-IP-model}.
}
\label{fig:pomex-dir}
\end{figure}

Diffractive collisions, producing  hadronic-states $X$ including a
 $c\bar{c}$ pair, are
a particularly interesting component of diffractive $ep$ interactions. 
The  charm-quark mass provides a hard scale, 
ensuring the applicability of perturbative QCD even 
for small photon virtualities (photoproduction). 
At leading order (LO) of QCD, two types of  photoproduction processes 
can be distinguished: direct and resolved photon processes. 
Charm production mainly proceeds 
via direct photon reactions, 
in which the exchanged photon participates as a point-like particle, 
directly interacting with a gluon from the incoming proton 
(photon-gluon fusion, Fig.~\ref{fig:pomex-dir}). 
Thus, diffractive charm production is directly
sensitive to the gluon content of the diffractive exchange. 
In the resolved photon processes, the photon behaves
as a hadron-like source of partons, one of which interacts with 
 a parton from the initial proton. Further interactions  
between partons  from  the  photon and the proton 
may fill the rapidity gap, leading to a suppression of the observed
cross sections in diffractive photoproduction.
For example, an  eikonal model \cite{ABK-eikmod} predicts a 
cross-section suppression by about a factor of three for 
diffractive resolved photoproduction at HERA.
A similar mechanism was proposed to explain the rate 
 of hard diffractive events 
at the Tevatron, which is 
lower than the expectations based on 
the dPDFs measured at HERA \cite{TevatronJJ}. 

  This paper presents a study of  diffractive charm production, 
  \mbox{$ep \to eD^*X^{\prime}p$},  
  with  exchanged-photon virtuality $Q^2 < 1 \gev^2$.
The production of charm was tagged by identification 
of a $D^{*\pm}(2010)$ meson in the final state\footnote{From now on, 
the notation $D^*$  will be used for both $D^{*+}$ and $D^{*-}$.}. 
The measurement is based on a sample of 
events with  a large gap in pseudorapidity between the proton and 
the produced hadronic system.
Diffractive charm production was measured previously at HERA  
in deep inelastic scattering (DIS) for 
photon virtualities  
 above $1.5 \GeV^2$ \cite{DdDISH1,DdPhPH1,ZeusDdDIS,DdDISpap}.
Recently, the H1  Collaboration has reported a measurement of 
diffractive  charm  photoproduction 
with $Q^2<0.01 \gev^2$ \cite{DdPhPH1}. 
The measurement reported here is performed with about six 
times larger statistics and in a larger kinematic range 
 than the H1 results. 

%
%
\section{Experimental set-up}
\label{sec-expset}
%
%

This measurement is based on the data taken 
with the ZEUS detector at the $ep$ collider HERA in 1998--2000, 
 when electrons or positrons of  27.5 $\gev$
  ~were collided with protons of  920 $\gev$. 
The sample used for this study  corresponds 
to an integrated luminosity 
$ \mathcal{L} = 78.6 \pm$ 1.8 pb$^{-1}$ 
(13.6  pb$^{-1}$ and 65.1  pb$^{-1}$ 
for the $e^-p$ and $e^+p$ samples, 
respectively\footnote{From now on, the word 
``electron'' will be used as a generic term 
for both electrons and positrons.}).

A detailed description of the ZEUS detector can be found elsewhere 
\cite{zeus:1993:bluebook}. 
Only a brief outline of the detector components most relevant 
to this analysis is given here.
 
Charged particles are tracked in the central tracking detector 
(CTD) \cite{CTD}, which
operates in a magnetic field of 1.43 T, provided by a thin super-conducting 
coil. The CTD consists of 72 cylindrical drift chamber layers, arranged  
in 9 superlayers, covering the polar angle region%
\footnote{The ZEUS coordinate 
system is a right-handed Cartesian system, with the $Z$ axis pointing 
in the proton beam direction, referred to as the ``forward direction'', 
and the $X$ axis pointing left towards the centre of HERA. 
The coordinate origin is at the nominal interaction point. }
\mbox{$15^\circ<\theta <164^\circ$}.
The transverse-momentum resolution for full-length tracks
is $\sigma(p_T )/p_T = 0.0058p_T \oplus 0.0065 \oplus 0.0014/p_T$, 
with $p_T$ in $\GeV$.

The high-resolution uranium--scintillator calorimeter (CAL) \cite{CAL} 
consists of three parts:
the forward (FCAL), the barrel (BCAL) and the rear (RCAL) calorimeters. 
Each part
is subdivided transversely into towers and longitudinally into one 
electromagnetic section 
 and either one (in RCAL) or two (in BCAL and FCAL) hadronic sections. 
 The smallest subdivision of the calorimeter is called a cell. 
The CAL energy resolutions, as measured under test-beam conditions, are 
$\sigma(E)/E = 0.18/\sqrt{E}$ for electrons
and $\sigma(E)/E = 0.35/\sqrt{E}$ for hadrons, with $E$ in $\gev$. 

In 1998--2000, the forward plug calorimeter  
(FPC) \cite{FPC} was installed in the 20 $\times$ 20 cm$^2$
beam hole of the FCAL with a small hole of radius 3.15 cm in the centre 
to accommodate the beam pipe. The FPC increased the forward calorimetric 
coverage by about one unit in pseudorapidity to $\eta ~\leq~ 5$. 

The luminosity was measured from the rate of the bremsstrahlung process 
$ep \to e\gamma p$.  The bremsstrahlung photons were measured 
with a lead--scintillator calorimeter \cite{LUMI} 
placed in the HERA tunnel at $Z = -107\,$m.

%
\section{Kinematics and reconstruction of variables}
\label{sec-kinvar}
%
%

Diffractive photoproduction in $ep$ scattering (Fig.~\ref{fig:pomex-dir}),
\[
e(e) + p(p) \to e(e^{\prime }) + X(X) + p(p^{\prime }),
 \]
is described in terms of the four-momenta
$e, e^{\prime}$ of the beam and scattered electrons, 
$p, p^{\prime}$ of the beam and scattered protons  and 
 $X$ of the hadronic system.
The following kinematic variables are defined:
the   photon virtuality, \mbox{$Q^2=-q^2$}, where
\mbox{$q=e-e^{\prime}$}, 
the squared photon-proton centre-of-mass energy,
$W^2=(p+q)^2$, and 
the fraction of the electron energy transferred 
to the proton in its rest frame, \\
\[ y=\frac{p\cdot q}{p\cdot e}\simeq\frac{W^2}{2p\cdot e}. \]
The reaction can be considered to proceed through 
the interaction of the virtual photon with 
the diffractive exchange (Pomeron, $\pom$). 
 This process  is described by
the invariant mass, $M_X$, 
of the hadronic system $X$ 
and the fraction of the proton momentum
\[ x_{\pomsub} = \frac{(p-p^{\prime})\cdot q}{p\cdot q}\]
 carried  by the diffractive exchange. 

The variables $W,$ $M_X$ and $x_{\pomsub}$ were reconstructed 
from the hadronic final state, using a combination of track and 
calorimeter information that optimises the resolution of 
the reconstructed kinematic variables. The selected tracks and 
calorimeter clusters are referred to as  energy-flow objects (EFO) 
\cite{EFO}. The Jacquet--Blondel formula~\cite{JB}
\[ W_{\JB} = \sqrt{2E_p(E-P_Z)} \] 
was used to reconstruct $W,$ where $E_p$ is the proton beam energy and
\[ E-P_Z = \sum_i (E_i-P_{Z_i}) ~. \]
 The invariant mass of the diffractively produced system 
 was calculated from
\[  M_X^2 = \left(\sum_i  E_i   \right)^2 - \left(\sum_i P_{X_i}\right)^2
      - \left(\sum_i P_{Y_i}\right)^2 - \left(\sum_i P_{Z_i}\right)^2.  \]
The sums in the above equations 
run over the energies $E_i$ and momentum components   
$P_{X_i}$ $P_{Y_i}$ and $P_{Z_i}$ of all EFOs. 

The variable $x_{\pomsub}$ was reconstructed from 
 \[ x_{\pomsub} = \frac{M_X^2}{W^2}, \]
which  is derived neglecting the photon virtuality 
($Q^2 \simeq 0$ for the case of photoproduction), 
the square of the four-momentum transfer at the
 proton vertex ($t=-(p-p^{\prime})^2$), and the mass of the proton. 

In addition, the inelasticity $z(D^*)$ was defined as \\ 
\[ z(D^*) = \frac{p \cdot p(D^*)}{p\cdot q},  \]
where $p(D^*)$ is the four-momentum of the $D^*$ meson. 
In the proton rest frame,
 $z(D^*)$ is the fraction of the photon energy carried 
 by the $D^*$ meson.
This variable was reconstructed as \\
\[z(D^*) = \frac{(E-P_Z)_{D^*}}{(E-P_Z)}, \] 
where $(E-P_Z)_{D^*}$ was calculated 
using the energy and momentum
component $P_Z$  of the $D^*$ meson. 

The measured values of the variables $W,$ $z(D^*)$, $M_X$ and $x_{\pomsub}$ 
 were corrected for energy losses in the inactive 
material of the ZEUS detector and for the loss of any particle
down the beam pipe using Monte Carlo (MC) simulations. 
All variables were reconstructed with a resolution of better than 15\%
over the ranges considered.

%
%
\section{Theoretical predictions}
\subsection{Monte Carlo simulations}
\label{subsec-mcsim}
%

Monte Carlo  simulations were used  to calculate the 
acceptances,  to evaluate  correction factors for 
 the selection inefficiencies and resolutions 
of the ZEUS detector and to estimate the background. 

The MC generator {\sc Rapgap} 2.08/18 \cite{RapGap} was used 
to simulate diffractive photoproduction of $D^*$ mesons. 
The simulation was performed  in the framework of the 
resolved-Pomeron model \cite{IS-IP-model}. 
The cross section is  proportional to the diffractive proton structure 
function, $F^D_2$, which is parameterised by the product of 
the probability of the Pomeron emission 
(the so-called Pomeron flux factor) 
and the structure function of the Pomeron. 
The  parameterisation of the Pomeron  flux factor
by  Streng and  Berger \cite{FluxF} was used along with 
the Pomeron  structure function obtained by the H1 Collaboration 
(H1Fit2 LO)~\cite{IP-H1Fit2}.  The contribution of the sub-leading 
Regge trajectory (the Reggeon),  which is only significant for  
$x_{\pomsub}>0.01$, was  also included. 

The $ep$ interactions  were modelled  
using both direct and resolved photon processes. 
The MC resolved photon component, 
which amounts to about 35\% of the total sample, is dominated 
by heavy-flavour excitation, in which a charm quark 
 from the photon participates in the hard scattering.
To simulate  resolved photon processes,
the GRV-G-LO \cite{GRV-G-LO} set of photon parton densities was used. 
The  simulation of charm production was performed 
with leading-order matrix elements. Higher-order QCD effects were
approximated by parton showers, based on the 
leading logarithm (LL) DGLAP splitting functions \cite{DGLAPSpF}.
Contributions from bottom production  with subsequent 
 decay into a final state with $D^*$ were also simulated. 
The bottom contribution, as predicted by the MC calculations, 
is not sizeable in any part of the kinematic range and  
corresponds to  2-3\% of the total sample. 
The masses of the heavy quarks were set
to  $m_c = 1.5 \GeV$ for charm and $m_b = 4.75 \GeV$ for  bottom.

The  MC generators  {\sc Pythia} 6.156  \cite{LundJETPYT} and
 {\sc Herwig} 6.301 \cite{HERWIG} were used  to model the 
 non-diffractive photoproduction of the $D^*$ mesons. 
 The CTEQ5L parameterisation \cite{CTEQ5}
 was used in both generators  for the proton PDFs.

The hadronisation process
was simulated with the  Lund string model \cite{LundJETPYT}
 in  the {\sc Rapgap} and  {\sc Pythia} MCs, and according to
a cluster hadronisation model\cite{HRWhadro} in  {\sc Herwig}. 

The generated Monte Carlo events were 
passed through the standard  
 simulation of the ZEUS detector, based on 
 the {\sc Geant} 3.13 simulation program \cite{GEANT}, 
 and  through the ZEUS  trigger simulation package \cite{ZTrg}. 
 The simulated detector responses were then subjected
to the same reconstruction and analysis procedures as the data.
For the determination of the acceptance  and   correction factors,
the generated {\sc Rapgap} events were re-weighted in 
the variables $M_X$ and $z(D^*)$, and the generated {\sc Pythia} 
and {\sc Herwig} events were re-weighted in  the variables 
$p_T(D^*)$ and $\eta(D^*)$  to improve the description of 
the shapes of the measured distributions.

\subsection{NLO QCD calculations}
\label{subsec-NLOQCD}
%
%

The cross sections for diffractive photoproduction of $D^*$ mesons 
were calculated at the  next-to-leading order (NLO) in  $\alpha_s$, 
the strong coupling constant, using the fixed-flavour-number scheme, 
in which only light flavours are active in the PDFs and the heavy 
quarks are generated by the hard  interaction. 
The calculation was performed with the FMNR code 
in the double-differential mode \cite{FMNR, applNLOQCD}. 
The Weizs\"acker-Williams approximation \cite{WWA} was used 
to obtain the virtual photon spectrum for electroproduction 
with small photon virtualities. Diffractive PDFs  were used instead 
of  the conventional proton PDFs. The three sets of dPDFs used in the 
calculations were derived from NLO QCD DGLAP fits to the HERA data on 
diffractive deep inelastic scattering: the H1 2006 Fit A, Fit B 
\cite{H1Fit2006} and the ZEUS LPS+charm Fit \cite{ZEUSLPS} diffractive 
PDFs. In the ZEUS LPS+charm fit, the diffractive DIS data were combined 
with the results  on diffractive charm production in DIS \cite{DdDISpap} 
to better constrain the gluon contribution. The Reggeon contribution, 
which amounts to less than 2\% for $x_{\pomsub}=0.01$ and grows up to 
$\sim 15$\% at $x_{\pomsub}=0.035$, was not included. To account for 
the proton-dissociative contribution, present in the H1 2006 fits, 
the corresponding predictions were multiplied by the factor 0.81 
\cite{H1Fit2006}. 

The calculations were performed with $\alpha_s(M_Z)=0.118 \gev$ 
\cite{PDG2006} and $m_c=1.45 \gev$, the same values  used in the 
QCD fits to the HERA data. The fraction of charm quarks hadronising 
as $D^*$ mesons was set to $f(c \to D^*) = 0.238$ \cite{cFF}. 
The Peterson parameterisation \cite{cFFPeter} was used for the charm 
fragmentation with the Peterson parameter $\epsilon= 0.035$,  
obtained in an NLO fit \cite{nlofitcFF} to ARGUS data \cite{ARGUS}. 
 The central NLO  QCD predictions were obtained with 
 the renormalisation and factorisation scales set to 
$\mu_R = \mu_F = 
\mu \equiv \sqrt{m_c^2+0.5\cdot [p_T^2(c)+p_T^2(\bar c)]}$.
Here,  $p_T(c)$ and $p_T(\bar c)$ are the transverse momenta of 
the charm and anti-charm quarks. The uncertainties of the 
calculations ~were~ estimated by  varying the renormalisation 
and factorisation  scales  simultaneously with the charm mass 
 to $\mu_R=\mu_F=0.5 \cdot \mu$, $m_c=1.25$ GeV and  to
    $\mu_R=\mu_F=2 \cdot \mu$, $m_c=1.65$ GeV 
 and they were found to be of the
order of $^{+30}_{-70}\%$. Variations of the charm mass only resulted
in a $\pm 15\%$ uncertainty.    
 Uncertainties on the dPDFs were not included.
 
  The NLO predictions are given by the sum of
   point-like and hadron-like processes, the NLO analogues of the
   direct and resolved photon processes defined at LO.
In all NLO calculations, the ~AFG-G-HO~ parameterisation \cite{AFG} 
was taken for the  photon PDFs. The hadron-like processes, in 
which the photon behaves as a source of light partons, 
contribute about 10\% of the FMNR cross section. 
The dependence on the photon PDFs was checked by using 
the ~GRV-G-HO~ parameterisation \cite{GRV-G-HO}
and was found to be negligible.
  It should be noted that the NLO diagrams in which the photon
  splits into a low-mass pair of $c$ and $\bar c$ quarks, one of 
  which interacts  with a gluon from the proton, are considered 
  as point-like photon  processes in FMNR  while they are effectively 
  included in {\sc Rapgap} as  resolved-photon processes with 
  heavy-flavour excitation.
  
In the calculations of  the 
 inclusive $D^*$ photoproduction cross sections, 
 the CTEQ5M parameterisation \cite{CTEQ5} with the default value 
of the QCD parameter ($\Lambda^{(5)}_{\rm QCD}=226 \mev$) was taken 
for the  PDFs of the proton.

%
%
\section{Event selection and reconstruction of $D^{*\pm}$ mesons}
\label{sec-dselec}
\subsection{ Event selection}
\label{subsec-trgphp}
%
%
%
 The events were selected online with a three-level trigger  system
\cite{zeus:1993:bluebook,ZTrg}. At the first- and  second-level triggers, 
data from CAL and CTD  were used to select $ep$ collisions and to reject 
non-$ep$ backgrounds. At the third level, the full event information was 
available and at least one reconstructed $D^*$ candidate (see below) 
was required. The efficiency of the online  $D^*$ reconstruction,
relative to the efficiency of the offline reconstruction,
was above $95\%$.

Photoproduction events ~were selected offline by ~requiring that 
~no ~scattered electron was identified in the CAL~\cite{no-electron}.
After correcting ~for~ detector effects, ~the most important~ 
~of~ which ~were~ energy losses ~in~ the inactive material 
~in~ front of the CAL and ~particle losses in ~~the ~beam pipe 
~\cite{no-electron, ZEUS-D*-incl}, ~~events ~were selected in 
~the interval ~$130 < W < 300 \GeV$ ($0.17<y<0.89$). 
The lower limit was set by the trigger requirements,
while the upper limit was imposed to suppress any
remaining events from  deep inelastic scattering.  
Under these conditions, the photon virtuality is below $1 \GeV^2$.
The median $Q^{2}$ value was estimated from a
MC simulation to be about $3 \times 10^{-4} \GeV^2$.

%
\subsection{ Reconstruction and selection of $D^{*\pm}$(2010) mesons}
\label{subsec-dspm}
%

The $D^*(2010)$ mesons were reconstructed from the decay 
\mbox{$D^* \to (D^0\to K\pi)\pi_s$}
by means of the mass-difference method  using charged tracks
measured in the CTD.  The $\pi_s$ particle from the $D^*$ decay 
is known as the ``soft'' pion because its momentum value 
is limited by the small difference between the masses of the 
$D^*$ and $D^0$ mesons.
To ensure a good efficiency and a good momentum resolution,  
tracks were required to have $p_T > 0.12 \gev$ and 
to reach at least the third CTD superlayer.

To reconstruct a $D^*$ candidate \cite{ZDs1996},  
two tracks of opposite charges  
were combined into a  $(K\pi)$ pair forming a $D^0$ candidate.
As kaons and pions were not identified, the mass of a charged kaon
and a charged pion was assigned to each track in turn. 

Similarly, 
to form a ``right-charge'' track combination for a $D^*$ candidate, 
each $(K\pi)$ pair   was combined with  a third track ($\pi_s$), 
which had the charged-pion mass assigned and charge opposite ~to~ 
that~ of  the ~$K$~ meson in the ~$(K\pi)$~ pair. 
To reduce~ ~the combinatorial background, 
~the tracks for the above combinations~ ~were~ selected~ 
with transverse momenta ~as~ follows:
$p_T(K) > 0.5 \GeV$, ~$p_T(\pi) > 0.5 \GeV$ ~and~ 
$p_T(\pi_s) > 0.12\GeV$. ~The  $p_T(\pi_s)$ cut ~was~ raised 
~to~ ~$0.25\GeV$~ ~for~ a data sub-sample, corresponding ~to~ 
an integrated luminosity ~of~ 16.9 $\pm$ 0.4 $\pb^{-1}$, 
for which the reconstruction efficiency of low momentum tracks 
was smaller ~due~ to the operating conditions ~of~ the CTD \cite{watercor}.
  ~The $D^*$-meson candidates were accepted provided  
 the invariant-mass value $M(K\pi)$ was consistent with the nominal 
$M(D^0)$ mass  given by the PDG \cite{PDG2006}. 
To take the mass resolutions into account, 
the following requirements ~were~ applied, depending  
~on~ $p_T (D^*)$, the transverse momentum of the $D^*$ meson 
\cite{ThetaC}: \\ \\
\hspace*{2.0cm} $ 1.82  < M(K\pi) < 1.91 \gev $ \hspace*{0.5cm}
for\hspace*{1.55cm} $p_T (D^*) < 3.25 \gev $, \\
\hspace*{2.0cm} $ 1.81  < M(K\pi) < 1.92 \gev $ \hspace*{0.5cm}
for\hspace*{0.25cm} $ 3.25< p_T (D^*) < 5 \gev$, \\
\hspace*{2.0cm} $ 1.80  < M(K\pi) < 1.93 \gev $ \hspace*{0.5cm}
for\hspace*{0.75cm} $5 < p_T (D^*) < 8 \gev$ and  \\
\hspace*{2.0cm} $ 1.79  < M(K\pi) < 1.94  \gev $ \hspace*{0.5cm}
for\hspace*{1.55cm} $p_T (D^*) > 8 \gev$.  \\ 

To suppress the combinatorial background further, 
the transverse momentum of the $D^*$ candidates 
was required to exceed $1.9\gev $ and a cut on the ratio 
$p_T(D^*)/E_T^{\theta>10^{\circ}}>0.1$ was applied.
Here $E_T^{\theta>10^{\circ}}$ is the transverse energy measured 
in the CAL outside a cone of $\theta = 10^{\circ}$ around the forward 
direction. Monte Carlo studies showed
that such a cut 
removes a significant fraction of the background 
whilst preserving most of the produced $D^*$ mesons.
The measurements were restricted to the pseudorapidity
range \( \vert \eta(D^*)\vert <1.6 \),
where the CTD acceptance is high.  
A clear signal was observed in the resulting mass difference 
$\Delta M=M(K\pi\pi_s)-M(K\pi) $ distribution (not shown) 
at the nominal value.   

To determine the number of $D^*$ mesons 
in the signal range, $0.1435 <\Delta M<0.1475 \gev $, the combinatorial
background was ~modelled~ by ``wrong-charge''~
track combinations and subtracted, after ~normalisation to
the ~``right-charge'' distribution ~in~ the range $0.15  < \Delta M < 0.17$ \Gev.
A ``wrong-charge'' track combination for a $(K\pi)$ pair
 was defined as two tracks of the same charge with 
 a soft pion ($\pi_s$) of the opposite charge. This subtraction
 yielded a signal of 12482 $\pm$ 208 inclusive  $D^*$ mesons.

\subsection{Selection of diffractive events}
\label{subsec-lrgselec}
%

Diffractive events ~were~ identified ~by~ 
the presence of a large rapidity gap (LRG)
~between~ the beam pipe, through which the scattered proton
 escaped detection,  and the hadronic-system $X$ ~\cite{pl:b315:481}. 
The events with a LRG were selected by applying a cut  on 
 the pseudorapidity $\eta_{\rm max}$
of the most  forward EFO with an energy greater than 400\mev. 
 
\begin{figure}[hbtp]
\epsfysize=20cm
\vspace*{-1.0cm}
\centerline{\epsffile{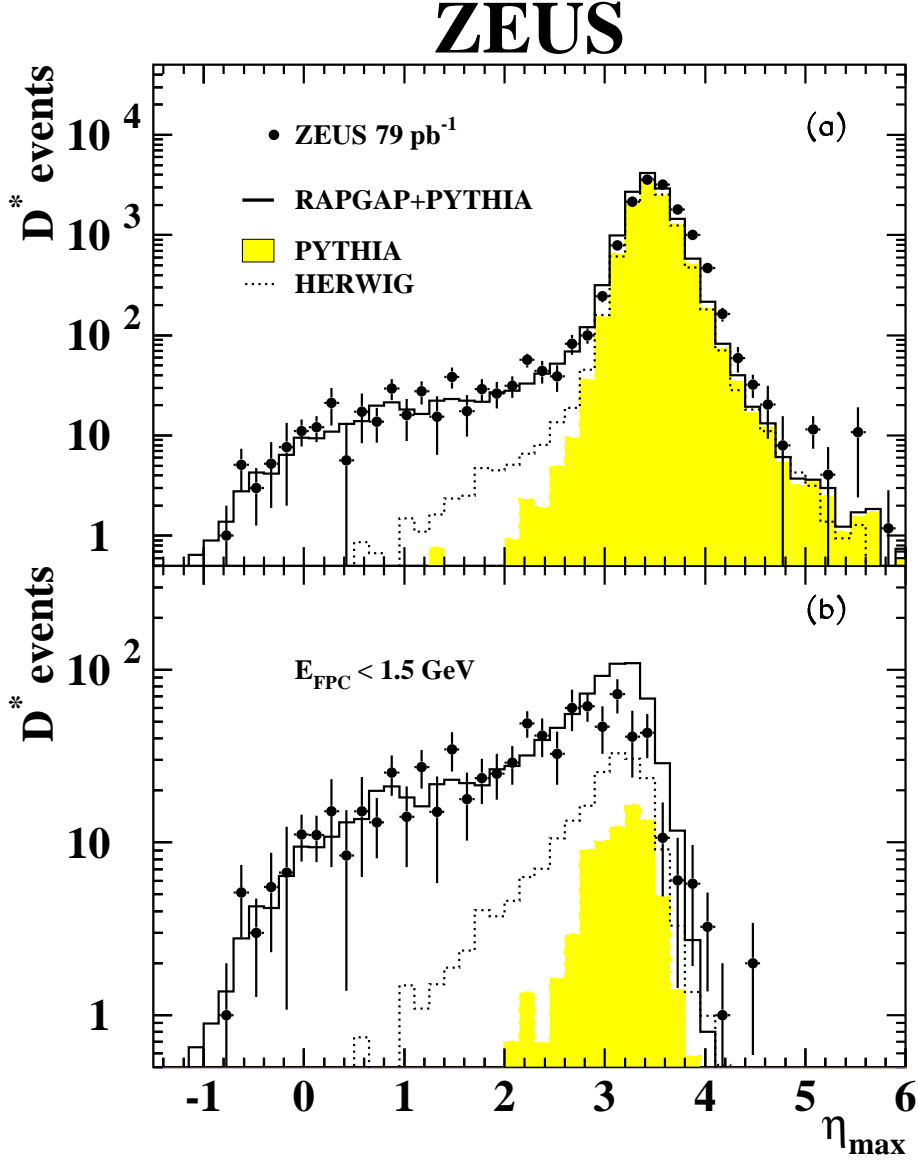}}
\vspace*{-2.0cm}
\caption{
Comparison of the measured $\eta_{max}$ distribution (dots)  
with the sum (solid histograms, normalised to the data) 
of the  ~weighted~ diffractive ({\sc Rapgap})  
~and~ non-diffractive ({\sc Pythia},  
shaded histograms) MC distributions for 
 (a) all inclusively photoproduced events with 
 a reconstructed $D^*$ meson  
and (b) events with $E_{FPC}<1.5 \gev$.
 The $D^*$ mesons with  
 $p_T(D^*)> 1.9 $ \Gev ~and \mbox{$\vert\eta(D^*)\vert< 1.6$}
 were selected in the kinematic region 
 $Q^2<1 \GeVs$ and $130<W<300\gev$.
The distributions for the non-diffractive events as predicted 
by {\sc Herwig} MC  are 
 indicated by the dotted histograms.
}
\label{fig:etamax}
\end{figure}

 Figure~\ref{fig:etamax}a compares  the measured $\eta_{\rm max}$ 
 distribution for all photoproduced $D^*$  events   
(after ``wrong-charge'' background subtraction) 
 to a sum of the distributions from diffractive ({\sc Rapgap}) 
and non-diffractive ({\sc Pythia})  MC samples.
The relative proportions of the two MC samples in the sum were chosen  
to give the best description of the shape of the data.  
The measured distribution shows two structures. The plateau 
at \mbox{$\eta_{\rm max} < 3 $} is populated predominantly 
by diffractive events, 
while the peak around $\eta_{\rm max} \sim 3.5 $ originates mainly from 
non-diffractive events. To select diffractive events, while rejecting the 
majority of the non-diffractive events,  
the energy deposited in the FPC  was required to be consistent with zero 
($E_{\rm FPC}<1.5 \gev$). Comparison between the  $\eta_{\rm max}$ 
 distributions of these  data and MC events 
 (Fig.~\ref{fig:etamax}b) confirms the considerable reduction 
 of non-diffractive events  in the sample. 
To further reduce the fraction of  non-diffractive events, 
a cut $\eta_{\rm max}< 3$ was applied.  
This cut ensures a gap of at least two units of pseudorapidity 
with respect to the edge of the forward calorimetric coverage 
provided by the FPC. A cut in $\eta_{\rm max}$
 correlates with the range of accessible $x_{\pomsub}$ values.
The requirement  $\eta_{\rm max}< 3$ restricts the measurement 
to \mbox{$x_{\pomsub} < 0.035$}.

\begin{figure}[hbtp]
\epsfysize=16cm
\vspace*{-2.0cm}
\centerline{\epsffile{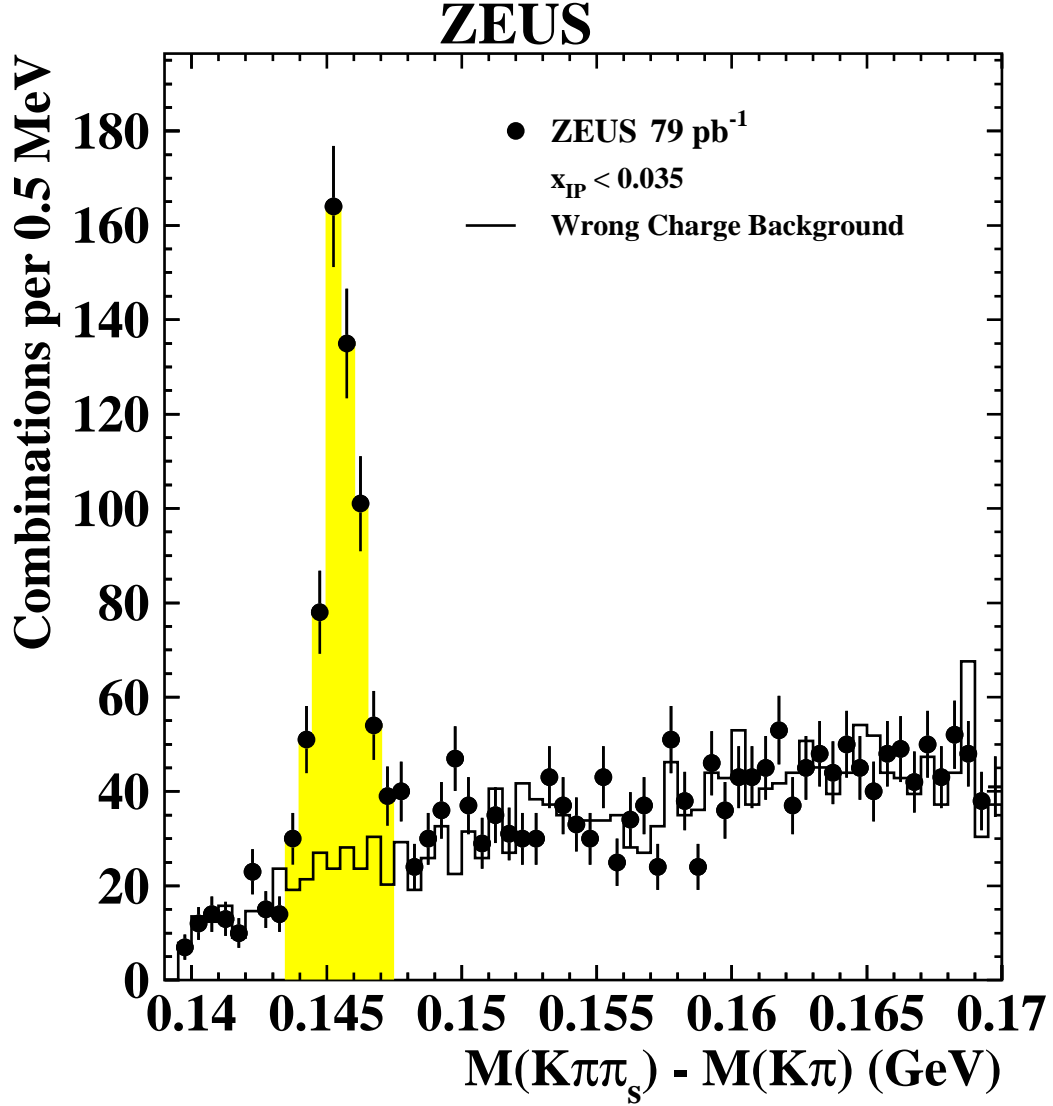}}
%
\caption{
The distribution of the mass difference,
$\Delta M=M(K \pi \pi_s)-M(K \pi)$, for the  
~$D^*(2010)$ candidates (dots)
with $p_T(D^*)>1.9 \gev$ and $\vert \eta(D^*) \vert<1.6$, 
reconstructed for $Q^2<1 \gev^2$, $130<W<300 \gev$ and $x_{\pomsub}<0.035$.
The shaded band shows the signal range, in which 
the combinatorial background (histogram) modelled by 
the wrong-charge combinations  was subtracted. 
 The signal contains  $458 \pm 30 ~D^*$ mesons.
}
\label{fig:ds035}
\end{figure}

%
After the above selections,   
 a signal of \mbox{458 $\pm$ 30}  $D^*$ mesons
was found in the $\Delta M$ distribution 
(Fig.~\ref{fig:ds035}) for diffractive photoproduction 
in the  range \mbox{$x_{\pomsub} < 0.035$}. 
In order to reduce the contributions from 
the Reggeon exchange and non-diffractive background,  
the selection was also performed in the restricted range 
{$x_{\pomsub} < 0.01$}, 
where~\mbox{204 $\pm$ 20 } $D^*$ mesons were observed.

%
%
From the comparison between the measured and MC 
$\eta_{\rm max}$ distributions (see above and Fig.~\ref{fig:etamax}a),  
normalisation factors were  obtained for 
the diffractive and the non-diffractive MC samples.  
These normalisation factors were then used to evaluate 
the  total  and differential fractions ($f_{\rm nd}$)
of residual non-diffractive events  in 
the  range \mbox{$\eta_{\rm max} < 3 $} and 
to correct all the measured distributions for this background bin-by-bin. 
The total fraction 
$f_{\rm nd}=3.3 \% $  was evaluated using the  {\sc Pythia} MC sample. 
Similar calculations were performed with the {\sc Herwig} MC sample 
  (total $f_{\rm nd}=15.5 \% $)
 for the purpose of systematic uncertainty evaluation.

The proton-dissociative events,  \mbox{$ep\to eXN$}, can also
satisfy the requirements $\eta_{\rm max}< 3$ and 
$E_{\rm FPC} < 1.5 \GeV$ if the proton-dissociative system, $N$,
has an invariant mass small enough to  pass undetected through 
the forward beam-pipe. The fraction ($f_{\rm pd}$) of background 
proton-dissociative events was measured
 previously to be $f_{\rm pd}=16 \pm 4(\rm syst.)$\%  \cite{DdDISpap},
 where the quoted uncertainty is due to the modelling 
and extraction procedure of the proton dissociation contribution. 
The proton dissociation contribution was assumed to be independent 
of all kinematic variables and was subtracted from all 
measured cross sections.

%
\section{Systematic uncertainties}
\label{sec-syserr}
%
%

The cross section uncertainties for  
$x_{\pomsub}<0.035$ and $~x_{\pomsub}<0.01$ were determined separately. 
The following sources of systematic  uncertainty were taken into account 
(uncertainties for the range $~x_{\pomsub}<0.01$ are given in brackets):

\begin{itemize}

 \item[$\bullet$] 
 the CAL simulation uncertainty was determined by 
 varying  the CAL energy scale by $\pm 2\%$ and
the CAL energy resolution  by $\pm 20\%$ of its value.
The CAL first-level trigger efficiencies were varied
 by their uncertainty. These variations resulted in a
combined $ ^{+1.8}_{-1.5}$ ($\pm 2.3$)\%  
uncertainty on the cross section;

 \item[$\bullet$] 
%
 the  tracking-simulation uncertainties were obtained by 
varying all momenta by $\pm 0.3$\% (magnetic field uncertainty) 
and by changing the track momentum and
angular resolutions by $^{+20}_{-10}$\% of their values. 
 The systematic uncertainty due to the simulation of the track inefficiency
 \cite{cFragmRatio} was found to be negligible.
 The variations  ~yielded~ a combined ~cross-section~ uncertainty  ~of~
  $^{+3.5}_{-1.9}$ ($^{+3.2}_{-3.3}$)\%;

 \item[$\bullet$]  
  the uncertainty in the subtraction of the combinatorial background
  was estimated ~by~ tightening separately ~by~ 2 MeV~ the lower ~and~ the
  upper boundary of the region in which the  ``wrong-charge'' background
  ~was~ normalised. ~This ~contributed \\ $+0.2$ ($-0.5$)\% to the
  cross-section uncertainty;

 \item[$\bullet$] 
the uncertainty in the FPC energy scale, evaluated by  $\pm 10$\%
 variations of  the FPC energy  in the MC,  
 gave a  systematic uncertainty of  $ ^{+0.2}_{-0.4}$ ($-0.2$)\%;  

 \item[$\bullet$]  
 the uncertainty in the selection of diffractive events 
was evaluated by varying  the  $E_{\rm FPC}$ cut by  $\pm 0.5$ GeV, 
which  yielded a cross-section 
 uncertainty of $ ^{+0.0}_{-0.9}$ ($^{+0.2}_{-0.3}$)\%, and 
  the  $\eta_{\rm max}$ cut  by  $\pm 0.2$, which 
  yielded a cross-section 
 uncertainty of $^{+6.3}_{-1.9}$ ($^{+2.6}_{-0.0}$)\%.
 The resulting uncertainty on  the selection of diffractive events 
 was  $^{+6.3}_{-2.1}$ ($^{+2.6}_{-0.3}$)\%;   

\item[$\bullet$] 
the uncertainty from the model dependence of the acceptance corrections 
  was determined by varying the re-weighting factors 
  of the  MC samples  by $\pm 20\%$ of their values.
The resulting cross-section uncertainty was 
   $^{+1.5}_{-1.4}$ ($^{+3.2}_{-3.3}$)\% ;

\item[$\bullet$] 
the uncertainty from the model dependence 
of  the non-diffractive event rejection 
 was determined using {\sc Herwig} instead of {\sc Pythia},  
  yielding a cross-section variation of $-11.9$ ($-6.8$)\%.
\end{itemize} 

The above systematic uncertainties were added in quadrature to 
determine the total systematic uncertainty.

%
The overall normalisation uncertainties due to 
the luminosity measurement ($\pm$2.2\%) and
 the $D^*$ ~and $D0$ decay  branching ratios ($\pm$2\%)
 were  not included in the total systematic uncertainty.
The cross section  uncertainty  due to 
  the subtraction of the proton dissociation  ($\pm$4.8\%) is 
  given separately.

%
\section{Results}
\label{sec-results}
\subsection{Cross sections}
\label{subsec-Csects}
%
%

The differential cross section for $ep \to eD^*X^{\prime}p$ 
in a given variable $\xi$ was calculated from
\[ 
\frac{d\sigma}{d\xi}=\frac{N_{D^*}\cdot(1-f_{\rm nd})\cdot(1-f_{\rm pd})}
 {\mathcal{A} \cdot \mathcal{L}\cdot \mathcal{B} \cdot\Delta\xi}~,
 \]
where ~$N_{D^*}$~ is the number of
~ $D^*$~ mesons observed in a bin of size ~$\Delta\xi$. 
 The overall acceptance was  ~$\mathcal{A}=13.9\%$.
 The combined
~\mbox{\( D^* \to (D^0 \to K \pi)\,\pi_s \)}~ decay branching ratio
 ~is ~\( \mathcal{B}=0.0257 \pm 0.0005\) \cite{PDG2006}.

The cross sections  for diffractive \(D^* \)-meson photoproduction 
were measured in the kinematic range 
  ~\mbox{$Q^2<1\gev^2$}, ~\mbox{$130 < W < 300\gev$} ($0.17<y<0.89$),
 ~\mbox{$ p_T(D^*)>1.9\gev$}, ~\mbox{$ \vert \eta(D^*)\vert < 1.6$}
 and ~\mbox{$ x_{\pomsub} < 0.035$}. 
 No restriction in $t$ was applied. The cross section,  
 integrated over this range, is  
\[ \sigma_{ep\to  eD^* X^{\prime}p}(x_{\pomsub}<0.035 )=
1.49 \pm 0.11(\rm stat.)^{+0.11}_{-0.19}(\rm syst.) \pm 0.07(p.d.)~\nb. \]

The last uncertainty  is due to  the subtraction of the 
proton-dissociative background (see Section \ref{subsec-lrgselec}).

The measurement was also repeated in the narrower range 
\mbox{$ x_{\pomsub} < 0.01$}, where the non-diffractive background 
admixture is smaller and the Reggeon contribution is expected to be 
negligible. The cross section integrated over the above kinematic 
region but for  \mbox{$ x_{\pomsub} < 0.01$} is   
\[ \sigma_{ep\to eD^* X^{\prime}p}(x_{\pomsub}<0.01)=
0.63 \pm 0.07(\rm stat.)^{+0.04}_{-0.06}(\rm syst.) \pm 0.03(p.d.)~\nb. \]

\begin{figure}[!h]
\vspace*{-4.0cm}
\hspace*{-3cm}\epsffile{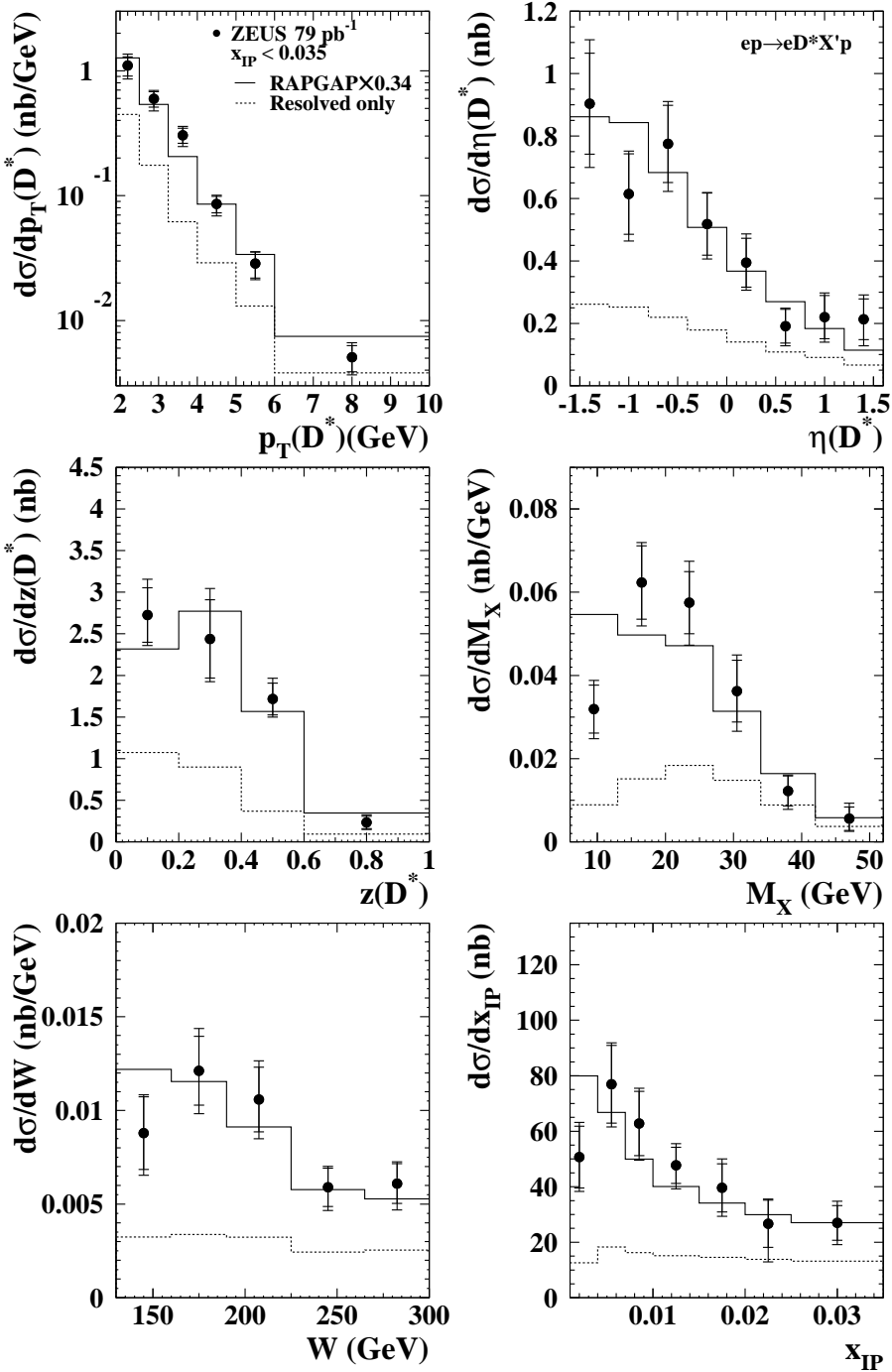}
\vspace*{-7.4cm}                                             
\caption{
Differential cross sections (dots) for  
diffractive photoproduction of $D^*$ mesons with respect to
 ~$p_T(D^*)$, ~$\eta(D^*)$, ~$z(D^*)$, ~$M_X$, ~$W$ 
 and ~$x_{\pomsub}$  measured  for ~$x_{\pomsub}<0.035$.
The inner bars show the statistical errors;  the outer bars 
correspond to the statistical and systematic uncertainties 
added in quadrature. The data are compared to the prediction 
of {\sc Rapgap} (solid histograms) using the  H1Fit2 LO  
diffractive parton   distribution parameterisation.  
The theoretical prediction  was normalised  to the data. 
The dashed histograms show the predicted contribution from resolved 
photon processes.  
}
\label{fig:cs035h1f2ll}
\end{figure}

\begin{figure}[hbtp]
\epsfysize=18cm
\vspace*{-2.0cm}
\hspace*{-1.0cm}
\epsffile{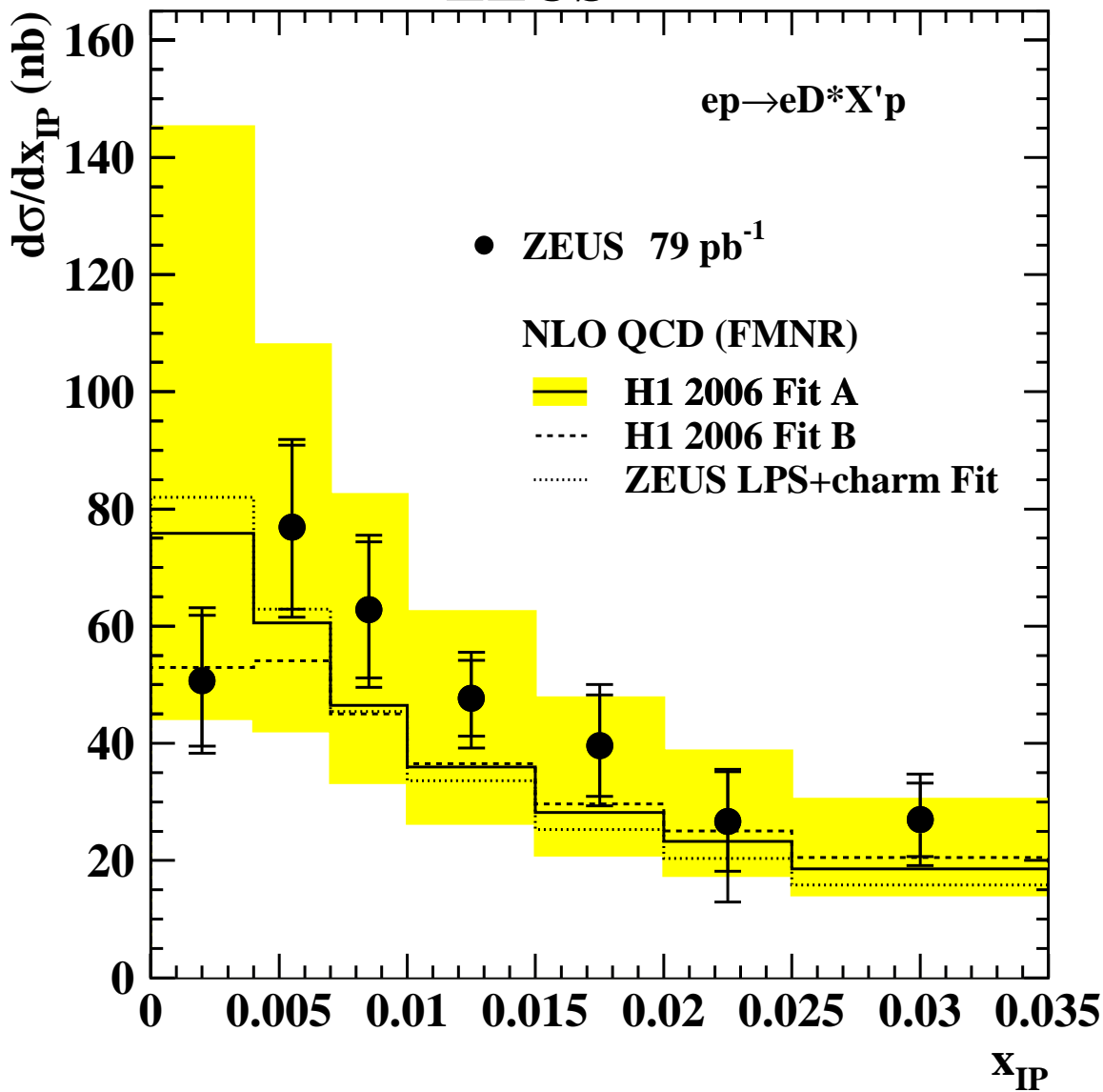}
\vspace*{-0.5cm}
\caption{
Differential cross section (dots) for 
diffractive photoproduction of $D^*$ mesons, 
measured  with respect to $x_{\pomsub}$.
The inner bars show the statistical errors; 
the outer bars correspond to the statistical 
and systematic uncertainties added in quadrature. 
 The data are compared to  the NLO QCD calculations (histograms) 
 using the H1 2006 Fit A (solid), Fit B (dashed), 
 both multiplied by a factor of 0.81, and
  the ZEUS LPS+charm Fit (dotted) diffractive 
 parton distribution parameterisations.  
The shaded bands show the uncertainties coming from variations 
of the charm-quark mass and the  
 factorisation and renormalisation scales. 
}
\label{fig:csx035zlpL243}
\end{figure}

\begin{figure}[!h]
\vspace*{-4.0cm}
\centerline{\epsffile{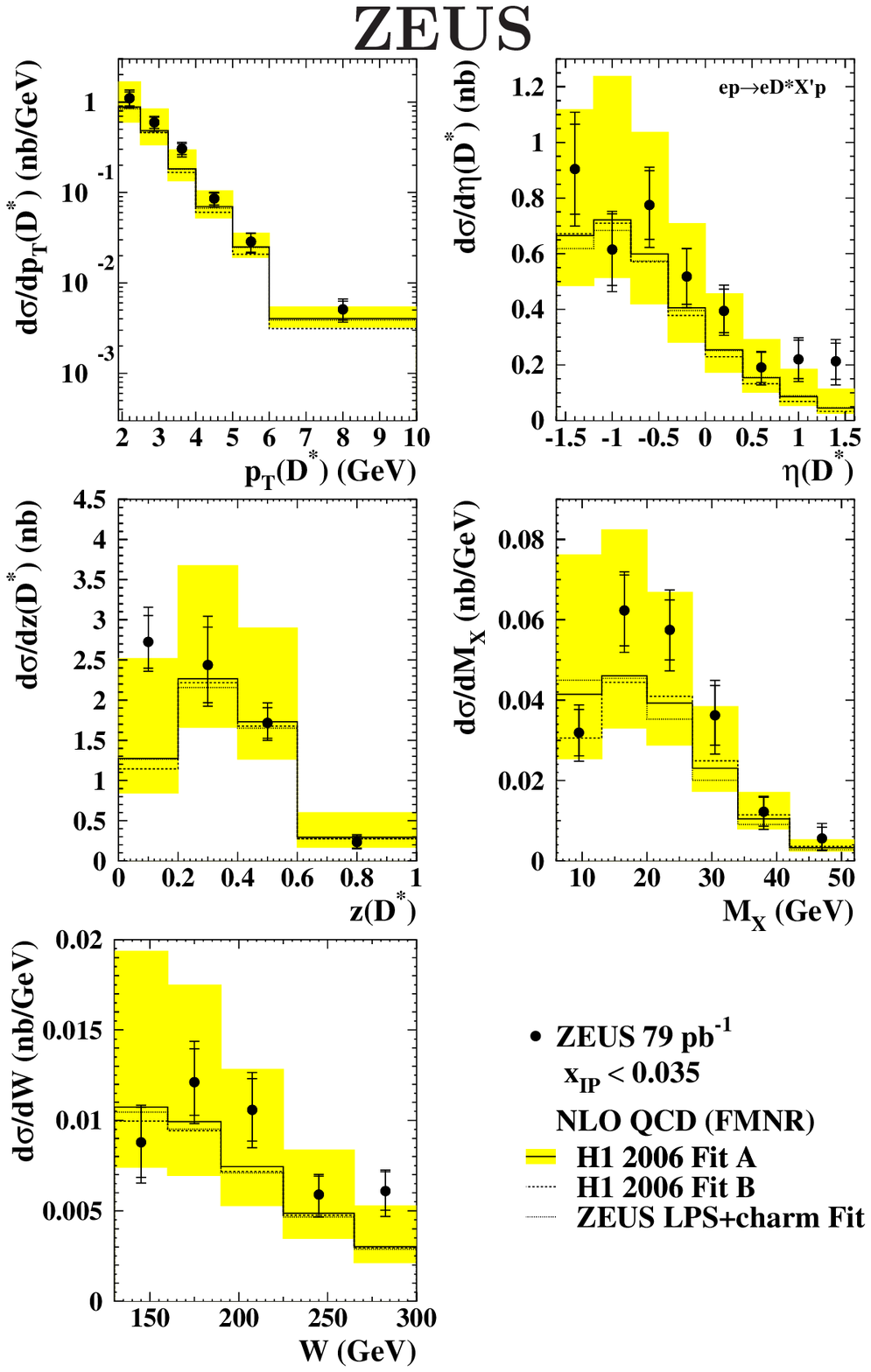}}
\vspace*{-7.7cm}                                             
\caption{
Differential cross sections (dots) for  
diffractive photoproduction of $D^*$ mesons with respect to 
 $p_T(D^*)$, $\eta(D^*)$, $z(D^*)$, $M_X$ and $W,$ 
 measured for $x_{\pomsub}<0.035$.
The inner bars show the statistical errors;  the outer bars 
correspond to the statistical and systematic uncertainties 
added in quadrature. 
 The data are compared to  the NLO QCD calculations (histograms) using 
 the  H1 2006 Fit A (solid), Fit B (dashed), 
 both multiplied by a factor of 0.81, and
 the  ZEUS LPS+charm Fit (dotted) 
 diffractive parton distribution parameterisations. 
The shaded bands show the uncertainties arising from  variations 
 of the charm-quark mass and the 
 factorisation and renormalisation scales.  
}
\label{fig:cs035zlpL243}
\end{figure}

\begin{figure}[!h]
\vspace*{-4.0cm}
\centerline{\epsffile{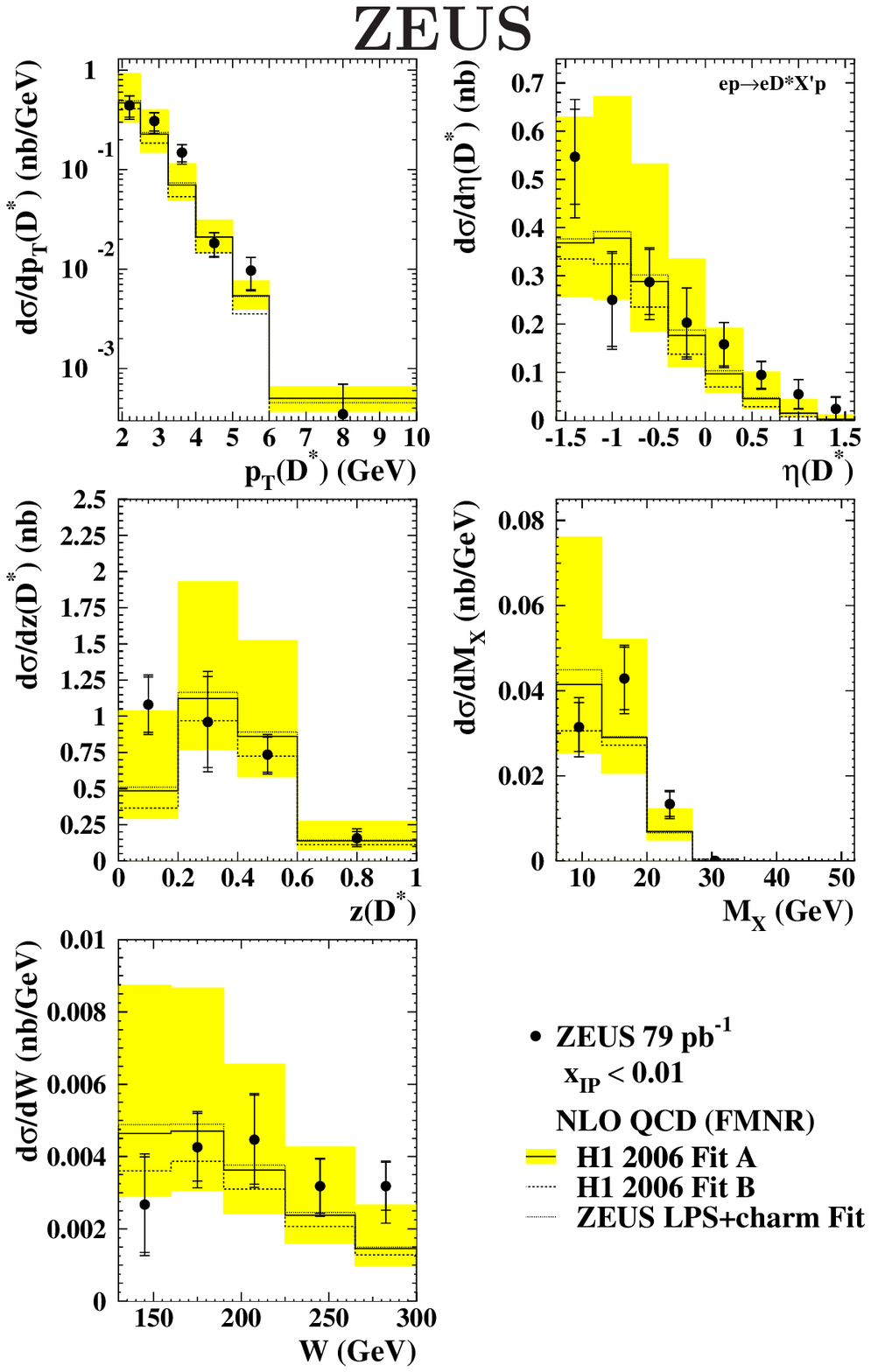}}
\vspace*{-7.7cm}                                             
\caption{
Differential cross sections (dots) for 
diffractive photoproduction of $D^*$ mesons with respect to 
 $p_T(D^*)$, $\eta(D^*)$, $z(D^*)$, $M_X$ and $W,$ 
 measured for $x_{\pomsub}<0.01$.
The inner bars show the statistical errors; the outer bars 
correspond to the statistical and systematic uncertainties added in quadrature. 
 The data are compared to  the NLO QCD calculations (histograms) using  
 the H1 2006 Fit A (solid), Fit B (dashed), 
 both multiplied by a factor of 0.81, and
   the  ZEUS LPS+charm Fit (dotted)  
 diffractive parton distribution parameterisations. 
The shaded bands show the uncertainties arising from variations 
of the charm-quark mass and the 
 factorisation and renormalisation scales. 
}
\label{fig:cs010zlpL243}
\end{figure}

For both $ x_{\pomsub}$ ranges, 
the differential cross sections, measured  as functions of the variables 
$x_{\pomsub}$, $M_X$, $p_T(D^*)$, $\eta(D^*)$, $z(D^*)$  and $W$,  
are presented in Tables \ref{tab:cspom}-\ref{tab:cswgp} and  
Figs.~\ref{fig:cs035h1f2ll}-\ref{fig:cs010zlpL243}. 

Figure~\ref{fig:cs035h1f2ll} ~compares~ the ~measured ~cross ~sections~~  
to~ the expectations ~from~ the resolved-Pomeron model calculated 
 by means of the ~{\sc Rapgap}~ MC program without re-weighting
 (Section~\ref{subsec-mcsim}).
To compare the shapes with the measured cross sections, 
the model prediction  was normalised by a factor 0.34. 
Reasonable agreement
between the shapes of the calculated and measured 
differential cross sections is observed. 
The relative contribution of resolved photon processes predicted by
{\sc Rapgap} increases  towards forward  $\eta(D^*)$, small   $z(D^*)$ 
and large  $M_X$. 

 Figures~\ref{fig:csx035zlpL243}-\ref{fig:cs010zlpL243}  
compare the measurements  to the three sets of NLO 
predictions obtained from the FMNR calculations 
using  the H1 2006 Fit A, Fit B and ZEUS LPS+charm Fit dPDFs.
 The estimated calculation uncertainties 
 (see Section \ref{subsec-NLOQCD}) 
 are shown  as the shaded band only for H1 2006 Fit A and are
 similar for other calculations.  
The uncertainties of the NLO QCD predictions are larger than
 the experimental ones in most bins.

The NLO QCD calculations  reproduce the   
$x_{\pomsub}$ differential cross section (Fig.~\ref{fig:csx035zlpL243}), 
in both shape and normalisation. A similar agreement  between
the calculations and the data is seen in 
Figs.~\ref{fig:cs035zlpL243} and \ref{fig:cs010zlpL243} for 
the $p_T(D^*)$, $\eta(D^*)$, $M_X$ and $W$ 
 differential cross sections in both ranges 
$ x_{\pomsub} < 0.035$ and $ x_{\pomsub} < 0.01$. 
The shapes of the differential distributions $d\sigma/dz(D^*)$   
 are not well reproduced by the NLO calculations.
A better shape description of 
the $z(D^*)$ distributions is provided 
by {\sc Rapgap} (Fig.~\ref{fig:cs035h1f2ll}).
The agreement between the  NLO QCD predictions and the data
supports the validity of the QCD factorisation theorem in diffraction, 
implying the universality of diffractive PDFs. 
 However, given the large experimental and theoretical
 uncertainties and the small hadron-like contribution expected
 by the NLO calculations, a suppression of the
 hadron-like component cannot be excluded.

\subsection{Fraction of $D^{*\pm}$ meson diffractive photoproduction }
\label{subsec-Ratio}
%
%

The fraction  of the diffractive to  the inclusive ($ep\to eD^*Y$) 
photoproduction cross sections for $D^*$ mesons was evaluated as 
\[
 \mathcal{R_D}(D^*) =
\frac{\sigma_{ep\to eD^* X^{\prime}p}(x_{\pomsub}<0.035
)}{\sigma_{ep\to eD^* Y}}.~\]

 In ~the~ ~kinematic region ~$Q^2 < 1 \GeV^2$, 
 ~$130 < W < 300 \GeV ~~(0.17<y<0.89)$,
~~$p_T(D^*)> 1.9  \GeV $ and $\vert\eta(D^*)\vert < 1.6$,  
diffractive production for  $x_{\pomsub}< 0.035$ contributes 

\[ \mathcal{R_D}(D^*) = 5.7 \pm
0.5(\rm stat.)^{+0.4}_{-0.7}(\rm syst.)  \pm 0.3(p.d.)\%~ \]
   
to the inclusive $D^*$ photoproduction  cross section. 
Systematic uncertainty partly cancel in this ratio.
The residual systematic
 uncertainty is dominated by the measurement of 
  the diffractive cross section. 
For the inclusive cross sections, the acceptance corrections were
estimated with {\sc Herwig}. The difference with respect to  
{\sc Pythia} was used as a systematic check.

\begin{figure}[!h]
\epsfysize=18cm
\vspace*{-1.0cm}
\centerline{\epsffile{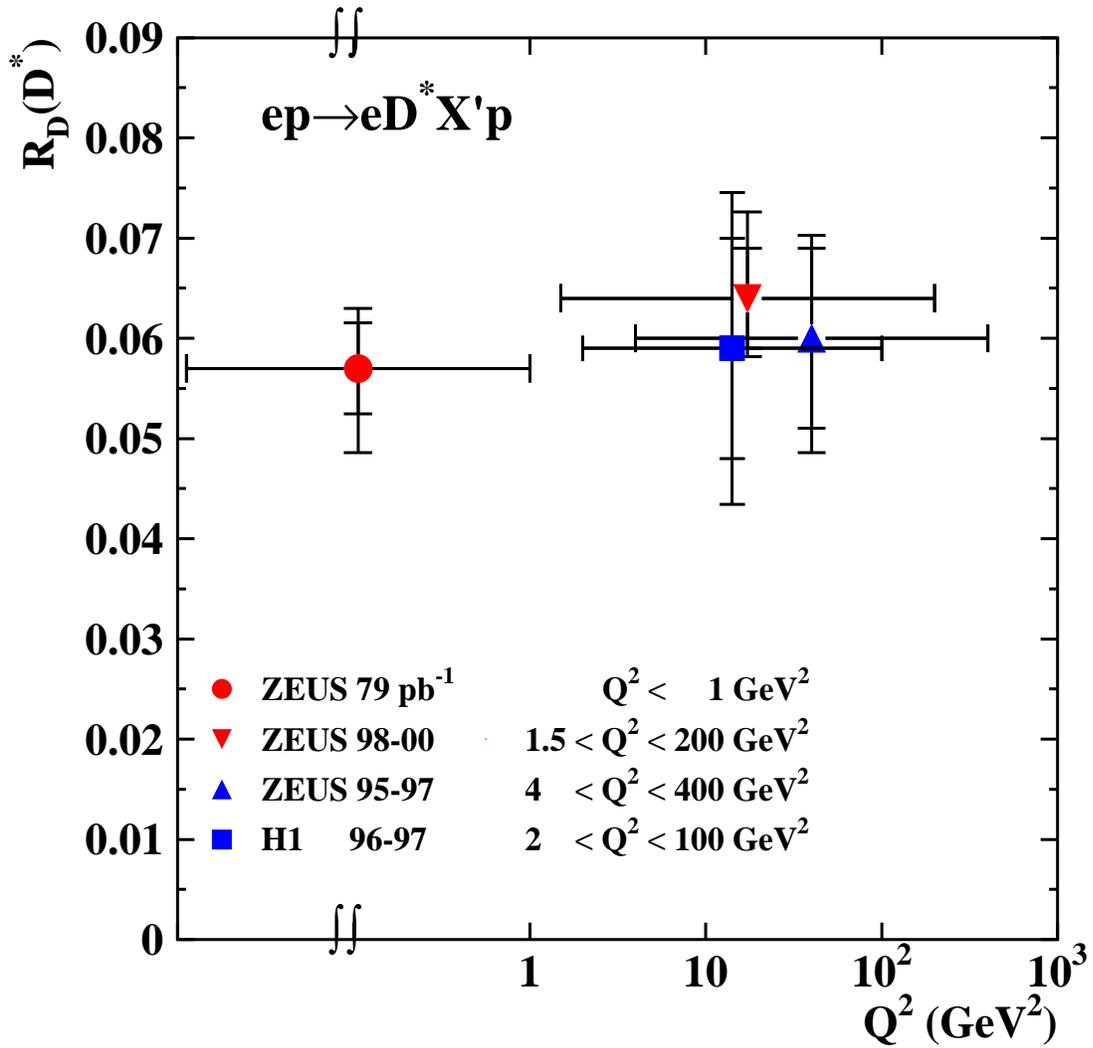}}
\vspace*{-2.5cm}
\caption{
Fractions  $\mathcal{R_D}$ of 
$D^*$ meson diffractive production cross sections                        
 measured at HERA  in DIS {\pcite{DdDISpap,ZeusDdDIS,DdDISH1}}  
 and  photoproduction  (this measurement).
The inner bars show the statistical errors, and the outer bars 
correspond to the statistical and systematic uncertainties 
added in quadrature. 
}
\label{fig:csRvsQ2}
\end{figure}

This  fraction $\mathcal{R_D}$   agrees with the values 
measured at HERA for diffractive DIS  in similar kinematic ranges
\cite{DdDISH1, DdDISpap, ZeusDdDIS}. As shown in Fig.~\ref{fig:csRvsQ2}, 
$\mathcal{R_D}$ is approximately independent of $Q^2$.

\begin{figure}[!h]
\vspace*{-0.55cm}
\epsfysize=30cm
\vspace*{-3.8cm}
\hspace*{-3.5cm}\epsffile{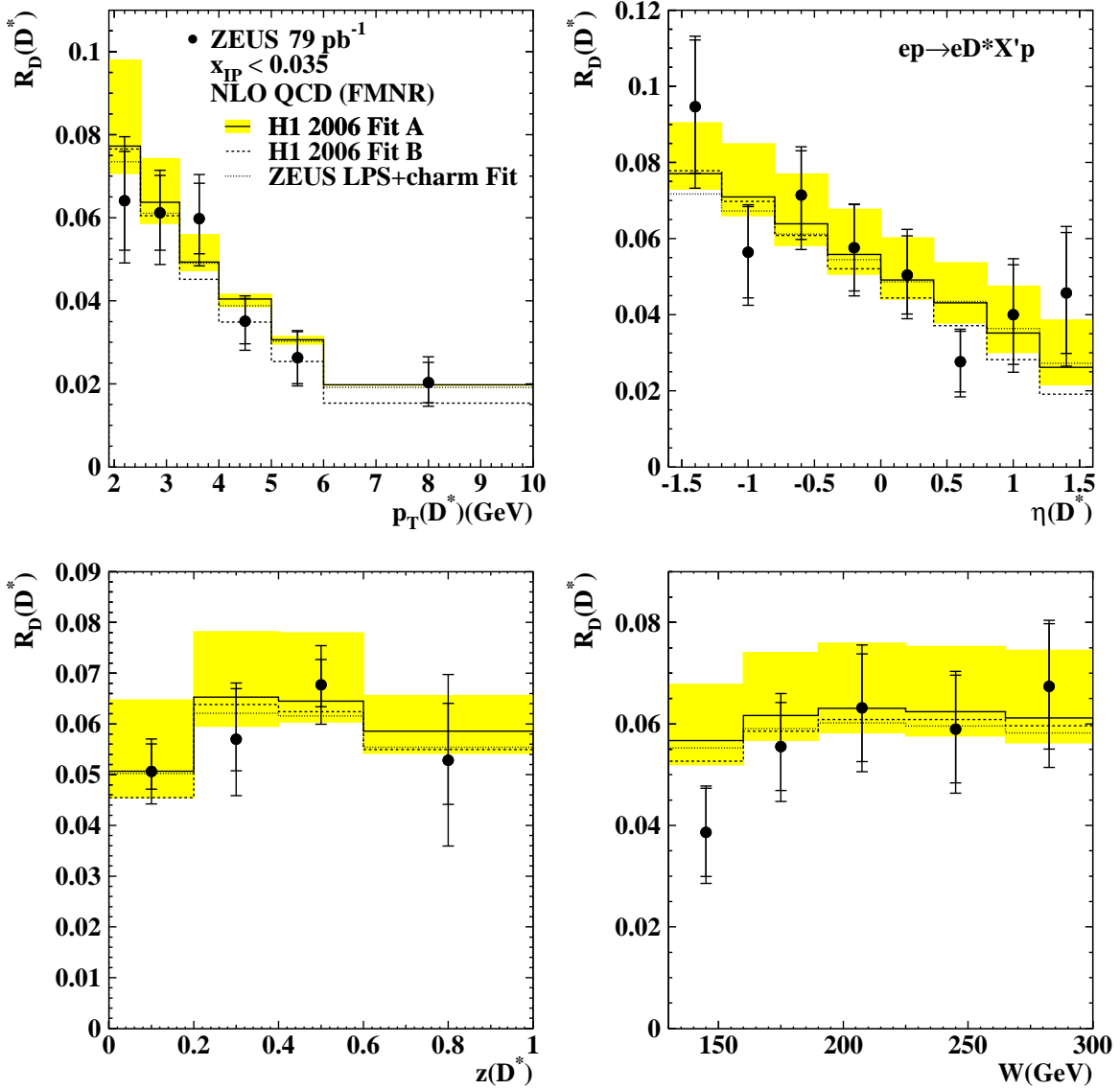}
\vspace*{-11.0cm}
\caption{
 Fraction of  $D^*$ meson diffractive photoproduction as a function of
 $p_T(D^*)$, $\eta(D^*)$, $z(D^*)$ and $W.$  The inner bars show 
the statistical errors;  the outer bars correspond to
 the statistical and systematic uncertainties added in quadrature.
 The data are compared to  the NLO QCD calculations (histograms) using
  the H1 2006 Fit A (solid), Fit B (dashed),
 both multiplied by a factor of 0.81, and
  the  ZEUS LPS+charm Fit (dotted)
 diffractive parton distribution parameterisations.
 The shaded bands show uncertainties arising from  variations
of the charm-quark mass and the
 factorisation and renormalisation scales.
}
\label{fig:csR035zlpL243}
\end{figure}

The differential dependences of the fraction $\mathcal{R_D}$ 
on  $p_T(D^*)$, $\eta(D^*)$, $z(D^*)$ and $W$ 
are shown in Tables \ref{tab:csd} and \ref{tab:cswgp} 
and Fig.~\ref{fig:csR035zlpL243}.
Similar to the  measurement in diffractive deep inelastic scattering 
\cite{DdDISpap}, the fraction of the diffractive contribution decreases
with  increasing   $p_T(D^*)$ and  $\eta(D^*)$.  
The value of $\mathcal{R_D}$ shows no strong dependence on 
either $W$ or  $z(D^*)$.

 The NLO QCD predictions for $\mathcal{R_D}$ were obtained as the
 ratio of the diffractive cross section, calculated
 with the H1 2006 or  ZEUS LPS+charm dPDFs, and the inclusive
 cross section, obtained with the  CTEQ5M proton PDFs.
The calculated ratios reproduce 
the measured dependence of $\mathcal{R_D}$ on the kinematic variables 
 well both in shape and normalisation 
(Fig.~\ref{fig:csR035zlpL243}), supporting diffractive QCD factorisation.

%
\section{Conclusions}
\label{sec-Summ}
%
%

Diffractive ~cross~ sections ~and~ their fraction ~of~ 
the total ~photoproduction~ cross ~section 
 of ~$D^{*\pm}(2010)$~  mesons have been measured with 
the ~ZEUS~ detector~ at~ HERA \\ 
~using~ an ~integrated luminosity  
of ~78.6 pb$^{-1}$. The ~$D^*$~ mesons ~were~ reconstructed with 
~$ p_T>1.9 \GeV$ ~and~ $\vert \eta\vert < 1.6$. 
~The measurements ~have~ been~ performed  
~in~ the  \\ kinematic region  ~$Q^2 < 1 \GeV^2$, 
~$130<W<300 \GeV~~(0.17 < y < 0.89)$, ~for~ the two ranges   
 $x_{\pomsub} < 0.035$ and $x_{\pomsub} < 0.01$.    

 The measured differential cross sections and 
 the fraction of the inclusive  photoproduction of
$D^{*\pm}$ mesons due to  diffractive exchange 
have been compared to the predictions of 
NLO QCD calculations using available parameterisations of
 diffractive PDFs. 
The NLO predictions based on H1 2006 fits A and B as well as
the ZEUS LPS+charm fit are consistent with the data.
The measured fraction of $D^{*\pm}$ meson photoproduction 
 due to  diffractive exchange 
is consistent with the measurements of $D^{*\pm}$ meson production
in diffractive  deep inelastic scattering.  
Within the experimental uncertainties, 
this fraction shows   no dependence on $Q^2$ and $W$.

 The results demonstrate that 
 diffractive open-charm photoproduction 
is well described by the   dPDF parameterisations extracted 
from diffractive DIS data, supporting the validity 
of  diffractive QCD factorisation. 
 However, given the large experimental and theoretical
 uncertainties and the small hadron-like contribution expected
 by the NLO calculations, a suppression of the
 hadron-like component  cannot be excluded.

%
\section{Acknowledgments}
\label{sec-Thanks}
We are grateful to the DESY Directorate for their strong support and
encouragement. The effort of the HERA machine group is gratefully
acknowledged. We thank the DESY computing and network services for their
support. The design, construction and installation of the ZEUS detector
have been made possible by the efforts of many people not listed as
authors.
It is a pleasure to thank H. Jung for useful discussions.
%
\vfill\eject

{
\def\bibname{\Large\bf References}
\def\refname{\Large\bf References}
\pagestyle{plain}
\ifzeusbst
  \bibliographystyle{./BiBTeX/bst/l4z_default}
\fi
\ifzdrftbst
  \bibliographystyle{./BiBTeX/bst/l4z_draft}
\fi
\ifzbstepj
  \bibliographystyle{./BiBTeX/bst/l4z_epj}
\fi
\ifzbstnp
  \bibliographystyle{./BiBTeX/bst/l4z_np}
\fi
\ifzbstpl
  \bibliographystyle{./BiBTeX/bst/l4z_pl}
\fi
{\raggedright
\bibliography{./BiBTeX/user/syn.bib,%
              ./BiBTeX/bib/l4z_articles.bib,%
              ./BiBTeX/bib/l4z_books.bib,%
              ./BiBTeX/bib/l4z_conferences.bib,%
              ./BiBTeX/bib/l4z_h1.bib,%
              ./BiBTeX/bib/l4z_misc.bib,%
              ./BiBTeX/bib/l4z_old.bib,%
              ./BiBTeX/bib/l4z_preprints.bib,%
              ./BiBTeX/bib/l4z_replaced.bib,%
              ./BiBTeX/bib/l4z_temporary.bib,%
              ./BiBTeX/bib/l4z_zeus.bib}}
}
\vfill\eject

%
%
\begin{table}[!thb]
\begin{center}
\begin{tabular}{||c||c||}\hline \hline
$x_{\pomsub}$  &  $d\sigma/dx_{\pomsub} ~(\nb)$   \\ 
 \hline \hline
0.0~~~~$\div$~0.004 &  $ 51~~ \pm 11~~^{~+~~6~~}_{~-~~5~~}$	\\
\hline
0.004~$\div$~0.007  &  $ 77~~ \pm 14~~^{~+~~5~~}_{~-~~6~~}$	\\
\hline
0.007~$\div$~0.010  &  $ 63~~ \pm 12~~^{~+~~5~~}_{~-~~6~~}$	\\
\hline
0.010~$\div$~0.015  &  $ 47.7 \pm ~6.5^{~+~~4.3}_{~-~~5.5}$	\\
\hline
0.015~$\div$~0.020  &  $ 39.6 \pm ~8.7^{~+~~5.8}_{~-~~5.5}$	\\
\hline
0.020~$\div$~0.025  &  $ 26.7 \pm ~8.5^{~+~~2.6}_{~-~10.8}$     \\
\hline
0.025~$\div$~0.035  &  $ 27.0 \pm ~6.3^{~+~~4.7}_{~-~~4.7}$     \\
 \hline \hline 
\end{tabular}
\caption{
Differential cross section  for 
 diffractive photoproduction of $D^*$ mesons 
as a function of $x_{\pomsub}$.
The first column shows the bin ranges. 
The first and the second uncertainties are 
respectively statistical and systematic. 
The overall normalisation uncertainties due to 
the luminosity measurement (2.2\%),  
the $D^*$ and $D^0$ branching ratios (2\%) and  
the proton-dissociative  contribution subtraction (4.8\%) 
are not indicated.
\label{tab:cspom}
%
         }
\end{center}
\end{table}
%
\begin{table}[!thb]
\begin{center}
\begin{tabular}{||c||c|c||}\hline \hline
$M_X$  & \multicolumn{2}{c||}{ $d\sigma/dM_X ~(\pb/\gev)$}   \\ 
 \hline
$(\gev)$ & $x_{\pomsub}<0.010$ & $x_{\pomsub}<0.035$    \\
 \hline \hline
~6~$\div$~13 & $  31.5 \pm 5.7^{~+~3.8}_{~-~4.1}$ &  
               $  31.9 \pm 5.8^{~+~3.8}_{~-~4.1}$   \\
\hline
13~$\div$~20 & $  42.9 \pm 7.4^{~+~2.5}_{~-~3.8}$  & 
               $  62.3 \pm 8.8^{~+~3.8}_{~-~5.6}$   \\
\hline
20~$\div$~27 & $  13.4 \pm 2.9^{~+~1.3}_{~-~1.8}$ & 
               $  57.5 \pm 7.5^{~+~6.6}_{~-~7.0}$   \\
\hline
27~$\div$~34 & $ 0.04 \pm 0.74^{+0.004}_{-0.006}$ & 
               $  36.2 \pm 7.4^{~+~4.5}_{~-~6.2}$   \\
\hline
34~$\div$~42 &                 $-$                &
               $  12.2 \pm 3.6^{~+~1.4}_{~-~2.6}$    \\
\hline
42~$\div$~52 &                 $-$                & 
               $  ~5.6 \pm 2.8^{~+~2.4}_{~-~1.1}$  \\
\hline \hline 
\end{tabular}
\caption{
Differential cross sections for  
diffractive photoproduction of $D^*$ mesons  
as a function of $M_X$ for the two ranges   
 $x_{\pomsub}<0.035$ and $x_{\pomsub}<0.01$.
The first column shows the bin ranges. 
The first and second uncertainties are 
respectively statistical and systematic. 
The overall normalisation uncertainties due to 
the luminosity measurement (2.2\%),  
the $D^*$ and $D^0$ branching ratios (2\%) and  
the proton-dissociative  contribution subtraction (4.8\%) 
are not indicated.
\label{tab:csmx}
%
         }
\end{center}
\end{table}
%
\begin{table}[!thb]
\begin{center}
\begin{tabular}{||c||c|c||c||}\hline \hline
%
%
$p_T(D^*)$  & \multicolumn{2}{c||}{ $d\sigma/dp_T(D^*)~(\pb/\gev)$} &
                                 $\mathcal{R_D}(p_T(D^*))$  \\ 
 \hline
$(\gev)$ & $x_{\pomsub}<0.010$ & $x_{\pomsub}<0.035$ &  (\%)  \\
 \hline \hline
1.9~~ $\div$ 2.5~  & $ 443 \pm 105^{~+~37}_{~-~60}$ 
            	  & $ 1100 \pm ~194^{~+~171}_{~-~145}$  
            	   	 & $ 6.4~\pm 1.2^{~+~1.0}_{~-~0.9}$  \\
\hline
2.5~~ $\div$ 3.25  & $ 308 \pm ~63^{~+~23}_{~-~45}$ 
            	  & $ ~~596 \pm ~~85^{~+~~52}_{~-~~84}$  
            	   	 & $6.1~\pm 0.9^{~+~0.5}_{~-~0.9}$  \\
\hline
3.25~$\div$ 4.0~  & $ 149 \pm ~29^{~+~~8}_{~-~19}$  
            	  & $ ~~304 \pm ~~42^{~+~~34}_{~-~~39}$ 
            	   	 & $ 6.0~\pm 0.8^{~+~0.6}_{~-~0.8}$  \\
\hline
4.0~ $\div$ 5.0~  & $ 18.3 \pm 4.9^{~+~1.2}_{~-~1.8}$      
            	  & $ ~~85.8 \pm13.1^{~+~~7.2}_{~-~11.0}$  
            	   	 & $ 3.5~\pm 0.5^{~+~0.3}_{~-~0.4}$  \\
\hline
5.0~ $\div$ 6.0~  & $ ~9.6 \pm 3.4^{~+~0.4}_{~-~1.0}$      
            	  & $ ~~28.6 \pm ~6.7^{~+~~2.1}_{~-~~3.0}$
                         & $ 2.6~\pm 0.6^{~+~0.2}_{~-~0.3}$  \\
\hline
6.0~ $\div$10.0~  & $ 0.35 \pm 0.35^{+0.03}_{-0.04}$ 
            	  & $ ~~5.09 \pm ~1.2^{~+~~1.0}_{~-~~0.8}$  
                         & $ 2.0~\pm 0.5^{~+~0.4}_{~-~0.3}$  \\
%
\hline \hline \hline
%
%
$\eta(D^*)$  & \multicolumn{2}{c||}{ $d\sigma/d\eta(D^*)~(\pb)$} &
                            $\mathcal{R_D}(\eta(D^*))$    \\ 
 \hline
 & $x_{\pomsub}<0.010$ & $x_{\pomsub}<0.035$ &  (\%)  \\
 \hline \hline
-1.6 $\div$ -1.2 & $ 547 \pm 98^{~+~66}_{~-~79}$  
                 & $ 904~\pm 162^{~+~125}_{~-~125}$  &
                       $ 9.5~\pm 1.8^{~+~0.6}_{~-~1.2}$  \\
\hline
-1.2 $\div$ -0.8 & $ 250 \pm 96^{~+~25}_{~-~35}$  
                 & $ 614~\pm 129^{~+~~48}_{~-~~78}$  &
                       $ 5.6~\pm 1.2^{~+~0.3}_{~-~0.7}$  \\
\hline
-0.8 $\div$ -0.4 & $ 287 \pm 68^{~+~21}_{~-~39}$  
                 & $ 775~\pm 124^{~+~~56}_{~-~~90}$  &
                       $ 7.1~\pm 1.1^{~+~0.5}_{~-~0.8}$ \\
\hline
-0.4 $\div$ ~0.0 & $ 203 \pm 71^{~+~10}_{~-~24}$ 
                 & $ 518~\pm 100^{~+~~18}_{~-~~51}$  &
                       $ 5.8~\pm 1.1^{~+~0.2}_{~-~0.6}$ \\
\hline
~0.0 $\div$ ~0.4   & $ ~158 \pm 45^{~+~~7.3}_{~-~18}$ 
                   & $ 394~\pm ~78^{~+~~49}_{~-~~40}$  &
                       $ 5.0~\pm 1.0^{~+~0.6}_{~-~0.5}$  \\
\hline
 ~0.4 $\div$  ~0.8 & $ ~~~95 \pm 27^{~+~~8.3}_{~-~11.8}$ 
                   & $ 191~\pm ~54^{~+~~20}_{~-~~32}$  &
                       $ 2.8~\pm 0.8^{~+~0.3}_{~-~0.5}$  \\
\hline
 ~0.8 $\div$  ~1.2 & $ ~~~55 \pm 30^{~+~~4.7}_{~-~~8.4}$ 
                   & $ 220~\pm ~69^{~+~~36}_{~-~~40}$  &
                       $ 4.0~\pm 1.3^{~+~0.7}_{~-~0.8}$  \\
\hline
 ~1.2 $\div$  ~1.6 & $ ~~~24 \pm 24^{~+~~8.6}_{~-~~8.9}$ 
                   & $ 213~\pm ~65^{~+~~43}_{~-~~55}$ &
                       $ 4.6~\pm 1.6^{~+~0.7}_{~-~1.1}$  \\
\hline \hline \hline
%
$z(D^*)$  & \multicolumn{2}{c||}{ $d\sigma/dz(D^*)~(\pb)$} &
                                  $\mathcal{R_D}(z(D^*))$  \\ 
 \hline
 & $x_{\pomsub}<0.010$ & $x_{\pomsub}<0.035$ &  (\%)  \\
 \hline \hline
0.0~$\div$~0.2 & $ 1080 \pm 191^{~+~~74}_{~-~~79}$ 
               & $ 2726 \pm 328^{~+~279}_{~-~166}$ & 
           	   $ 5.1 \pm 0.6^{~+~0.5}_{~-~0.4}$   \\
\hline
0.2~$\div$~0.4 & $ ~960 \pm 315^{~+~152}_{~-~137}$ 
               & $ 2438 \pm 470^{~+~384}_{~-~207}$ & 
           	   $ 5.7 \pm 1.1^{~+~1.0}_{~-~0.6}$   \\
\hline
0.4~$\div$~0.6 & $ ~735 \pm 121^{~+~~67}_{~-~~59}$ 
               & $ 1717 \pm 190^{~+~160}_{~-~107}$ & 
           	   $ 6.8 \pm 0.8^{~+~0.5}_{~-~0.4}$   \\
\hline
0.6~$\div$~1.0 & $ ~157 \pm ~46^{~+~~45}_{~-~~36}$ 
               & $ ~234 \pm ~~74^{~+~~55}_{~-~~43}$ & 
                   $ 5.3 \pm 1.7^{~+~1.1}_{~-~0.9}$   \\
\hline \hline
\end{tabular}
\caption{
Differential cross sections for 
diffractive photoproduction of $D^*$ mesons  
for  the two  ranges  $x_{\pomsub}<0.035$ and $x_{\pomsub}<0.01$ and 
diffractive fraction $\mathcal{R_D}$ of $D^*$ meson photoproduction 
as  functions of $p_T(D^*)$, $\eta(D^*)$ and $z(D^*)$. 
The first column shows the bin ranges. 
The first and second uncertainties are  
respectively statistical and systematic. 
The overall normalisation  uncertainties due to 
the luminosity measurement (2.2\%),  
the $D^*$ and $D^0$ branching ratios (2\%) and  
the proton-dissociative  contribution subtraction (4.8\%) 
are not indicated.
\label{tab:csd}
%
         }
\end{center}
\end{table}
\vspace*{-25.5cm}
%
\newpage
\mbox{   } \\
\vspace*{6cm}  
%
\begin{table}[!thb]
\begin{center}
\begin{tabular}{||c||c|c||c||}\hline \hline
$W$  & \multicolumn{2}{c||}{ $d\sigma/dW~(\pb/\gev)$} &
                                  $\mathcal{R_D}(W)$  \\ 
 \hline
$(\gev)$ & $x_{\pomsub}<0.010$ & $x_{\pomsub}<0.035$ &  (\%)  \\
\hline \hline
130 $\div$ 160 & $ 2.7 \pm 1.3^{~+~0.5}_{~-~0.5}$ 
& $ ~8.8~\pm 1.9^{~+~0.7}_{~-~1.2}$  &  
                $ 3.9~\pm 0.9^{~+~0.3}_{~-~0.5}$  \\
\hline
160 $\div$ 190 & $ 4.3 \pm 0.9^{~+~0.3}_{~-~0.6}$ 
& $ 12.1~\pm 1.8^{~+~1.3}_{~-~1.4}$  &  
                $ 5.6~\pm 0.9^{~+~0.6}_{~-~0.7}$ \\
\hline
190 $\div$ 225 & $ 4.5 \pm 1.2^{~+~0.3}_{~-~0.5}$ 
& $ 10.6~\pm 1.7^{~+~1.1}_{~-~1.2}$  &  
                $ 6.3~\pm 1.1^{~+~0.6}_{~-~0.7}$ \\
\hline
225 $\div$ 265 & $ 3.2 \pm 0.7^{~+~0.2}_{~-~0.4}$ 
& $ ~5.9~\pm 1.0^{~+~0.5}_{~-~0.7}$  & 
                $ 5.9~\pm 1.1^{~+~0.4}_{~-~0.7}$  \\
\hline
265 $\div$ 300 & $ 3.2 \pm 0.7^{~+~0.2}_{~-~0.8}$ 
& $ ~6.1~\pm 1.1^{~+~0.5}_{~-~0.9}$  & 
                $ 6.7~\pm 1.2^{~+~0.4}_{~-~1.0}$  \\
 \hline \hline 
\end{tabular}
\caption{
Differential cross section  for  
diffractive photoproduction of $D^*$ mesons  
for the two ranges  $x_{\pomsub}<0.035$ and $x_{\pomsub}<0.01$ and   
diffractive fraction $\mathcal{R_D}$ of  
$D^*$ meson photoproduction  as a function of $W$.
The first column shows the bin ranges. 
The first and second uncertainties are respectively
statistical and systematic. The overall normalisation
uncertainties due to the luminosity measurement (2.2\%),  
the $D^*$ and $D^0$ branching ratios (2\%) and  
the proton-dissociative  contribution subtraction (4.8\%) 
are not indicated.
\label{tab:cswgp}
         }
\end{center}
\end{table}
\vspace*{-25.5cm}
%

%
\newpage
\mbox{   } \\
\vspace*{0.5cm}  
%
%
%
%
%
%
%
%
%
%
%
%
%
%
%
%
%
%
%
.................................................................

\end{document}